\documentclass[preprint,aps,floats,nofootinbib,amssymb,superscriptaddress]{revtex4}
\pdfoutput=1
\usepackage[sort&compress]{natbib}
\usepackage{epsf,epsfig}
\usepackage{amsmath}
\usepackage{subfigure}
\usepackage{graphicx}
\usepackage{color}
\usepackage{mathrsfs}
\usepackage[utf8]{inputenc}

\newcommand{\be}{\begin{equation}}
\newcommand{\ee}{\end{equation}}
\newcommand{\ba}{\begin{eqnarray}}
\newcommand{\ea}{\end{eqnarray}}
\newcommand{\nn}{\nonumber}

\newcommand{\bi}{\begin{itemize}}
\newcommand{\ei}{\end{itemize}}
\newcommand{\vf}{\sqrt{v_\phi^2+v^2 s^2_{2\beta}}}
\newcommand{\vvf}{v_\phi^2+v^2 s^2_{2\beta}}
\newcommand{\der}{\stackrel{\leftrightarrow}{\partial}}

\makeatletter

\begin{document}

\preprint{ICCUB-15-009}

\title{\vspace*{1.in} Axion-Higgs interplay in the two Higgs-doublet model}

\author{Dom\`enec Espriu}\affiliation{Departament d'Estructura i Constituents de la Mat\`eria}
\author{Federico Mescia}\affiliation{Departament de F\'\i sica Fonamental,
Institut de Ci\`encies del Cosmos (ICCUB), \\
Universitat de Barcelona, Mart\'i Franqu\`es 1, 08028 Barcelona, Spain}
\author{Albert Renau}\affiliation{Departament d'Estructura i Constituents de la Mat\`eria}

\vspace*{2cm}

\thispagestyle{empty}

\begin{abstract} 
The DFSZ model is a natural extension of the two-Higgs
doublet model containing an additional  singlet, endowed with a Peccei-Quinn symmetry, 
and leading to a physically acceptable  axion. In this paper we re-examine this model
in the light of some new developments. For generic couplings the model 
reproduces the minimal Standard Model showing only tiny deviations (extreme decoupling scenario) 
and all additional degrees of freedom (with the exception of the axion) are
very heavy. Recently it has been remarked that the limit where the coupling between the singlet
and the two doublets becomes very small is technically natural. Combining this 
limit with the requirement of exact or approximate custodial symmetry 
we may obtain an additional $0^+$ Higgs at the weak scale, accompanied by relatively light charged 
and neutral pseudoscalars. The mass spectrum would then resemble that of a generic two Higgs-doublet
model, with naturally adjustable masses in spite of the large scale that the axion introduces. 
However the couplings are non-generic in this model. We use the recent constraints
derived from the Higgs-WW coupling together with oblique corrections to constrain the model
as much as possible. As an additional result, we work out the non-linear parametrization 
of the DFSZ model in the generic case where all scalars  except the lightest Higgs and the 
axion have masses at or beyond the TeV scale.  
\end{abstract}

\maketitle

\section{Introduction}\label{sec:intro}
An invisible axion~\cite{PQ,sik,raff} constitutes to this date a very firm candidate to provide
all or part of the dark matter component of the cosmological budget. There are several 
extensions of the Minimal Standard Model (MSM) providing a particle with the characteristics
and couplings of the axion~\cite{dfs,models}. In our view a particularly interesting possibility 
is the model suggested by Zhitnitsky and Dine, Fischler and  Srednicki (DFSZ)
more than 30 years ago that consists in a fairly simple extension of the popular two Higgs-doublet model (2HDM). 
As a matter of fact it could probably be argued that a good motivation to consider the 2HDM is 
that it allows for the inclusion of
a (nearly) invisible axion~\cite{Axion,2hdmNatural, 2hdmConf}. 
Of course there are other reasons why the 2HDM should be considered
as a possible extension of the MSM. Apart from purely aestethic 
reasons, it is easy to reconcile such models with existing constraints. 
They may give rise to a rich phenomenology, including possibly
(but not necessarily) flavour changing neutral currents at some level, custodial 
symmetry breaking terms or even new sources of CP violation~\cite{2HDM,2HDM2}.

Following the discovery of a Higgs-like particle with $m_h\sim 125$ GeV there have been a number 
of works considering the implications of such a finding on a generic 2HDM, together with
the constraints arising from the lack of detection of new scalars and from the electroweak
precision observables~\cite{pich}. Depending on the way that the
two doublets couple to fermions, they are classified as type I, II or III (see e.g.~\cite{2HDM}
for details), with different implications on the flavour sector. Consideration of all the different 
types of 2HDM plus all the rich phenomenology that can be encoded in the Higgs potential leads
to a wide variety of possibilities with different experimental implications, 
even after applying all the known phenomenological low-energy requirements.

Requiring a Peccei-Quinn (PQ) symmetry leading to an axion does however severely restrict the
possibilities, and this is in our view an important asset of the DFSZ model.
This turns out to be particularly the case when one includes all the recent experimental 
information concerning the 125 GeV scalar state and its couplings. Exploring this model, taking
into account all these constraints is the motivation for the present work. 
 
The structure of this paper is as follows. In section \ref{sec:mas} we discuss 
the possible global symmetries of the DFSZ model, namely $U(1)_{PQ}$ (always present), 
and $SU(2)_L\times SU(2)_R$ (the $SU(2)_R$ subgroup may or may not be present). Symmetries are 
best discussed by using a matrix formalism that we review and extend. 
Section \ref{sec:mam} is devoted to the determination
of the spectrum of the theory. We present up to four generic cases that range from the extreme
decoupling, where the model --apart from the presence of the axion-- is indistinguishable from the 
MSM at low energies, to one where there are extra light Higgses below or around the TeV scale.
This last case necessarily requires some
couplings in the potential to be very small; a possibility that is nevertheless natural in
a technical sense and therefore should be contemplated as a viable theoretical hypothesis. 
We discuss in detail the situation where custodial symmetry is exact or approximately
valid in this model because the combination of this symmetry and naturally
small couplings allows us to keep the additional scalars `naturally light' if we so wish with
only one exception, meaning that the `contamination' from the large scale present in the theory is under 
control.   

However, while additional scalars may exist at or just above the weak scale in
this model, they can also be made heavy, with masses in the multi-TeV region or beyond.
In section   \ref{sec:nonlinear} we discuss the resulting non-linear effective theory 
emerging in this generic situation. 

Next in section \ref{sec:higgs} we analyze the impact of the model on the (light) 
Higgs effective couplings to gauge bosons and the constraints that can be derived 
from the recent LHC data. In section \ref{sec:matching} we compare the potential of the DFSZ model with the most general potential in the 2HDM. 
We find out which terms of the general potential are forbidden by the PQ symmetry and which ones are recovered when it is spontaneously broken by the VEV of 
the scalar fields.
Finally in section \ref{sec:deltarho} the restrictions 
that the electroweak precision parameters, particularly $\Delta\rho$, impose on the 
model are discussed. These restrictions are relevant only in the case where 
all or part of the additional spectrum of scalars is light as 
we find that they are automatically satisfied otherwise.

We would like to emphasize that even after imposing the constraints derived from the PQ symmetry 
the model still contains enough degrees of freedom to reproduce the mass spectrum of a generic
2HDM, so there is no predictivity at the level of the spectrum. However, this nearly exhausts all
freedom available, particularly if exact or approximate custodial symmetry is imposed. 
Then the scalar couplings are largely fixed and in this sense the model
is far more predictive than a generic 2HDM ---and in addition it contains the axion, which
is its {\em raison d'\^etre}.

\section{Model and symmetries}\label{sec:mas}
The DFSZ model contains two Higgs doublets and one complex scalar singlet, namely
\be
\phi_1=\left(
\begin{array}{c}
\alpha_+\\
\alpha_0
       \end{array}
       \right);\quad
\phi_2=\left(
\begin{array}{c}
\beta_+\\
\beta_0
       \end{array}
       \right);\quad
\phi,
\label{eq:fields}
\ee
with vacuum expectation values (VEVs) 
$\langle\alpha_0\rangle=v_1$, $\langle\beta_0\rangle=v_2$, 
$\langle\phi\rangle=v_\phi$  and $\langle\alpha_+\rangle=\langle\beta_+\rangle=0$.
Moreover, we define the usual electroweak vacuum
expectation value $v=246$ GeV as
$v^2=(v_1^2+v_2^2)/2$ and $\tan\beta=v_2/v_1$. The implementation of the PQ symmetry is only possible for type II models, where the
Yukawa terms are
\be\label{yuk}
\mathcal{L}_Y=G_1\bar q_L\tilde\phi_1u_R+G_2\bar q_L\phi_2d_R+h.c.,
\ee
with $\tilde\phi_i=i\tau_2\phi^*_i$. The PQ transformation acts on the scalars as
\be
\phi_1\to e^{iX_1\theta}\phi_1,\quad\phi_2\to e^{iX_2\theta}\phi_2,\quad\phi\to e^{iX_\phi\theta}\phi
\ee
and on the fermions as
\be
q_L\to q_L,\quad l_L\to l_L,\quad u_R\to e^{iX_u\theta}u_R,\quad d_R\to e^{iX_d\theta}d_R,\quad e_R\to e^{iX_e\theta}e_R.
\ee
For the Yukawa terms to be PQ-invariant we need
\be
X_u=X_1,\quad X_d=-X_2,\quad X_e=-X_2.
\ee
Let us now turn to the potential involving the two doublets and the new complex singlet. The most
general potential compatible with PQ symmetry is 
\ba\label{potential}
\!\!\!V(\phi,\phi_1,\phi_2)&=&\lambda_\phi(\phi^*\phi-V_\phi^2)^2+\lambda_1(\phi_1^\dag\phi_1-V_1^2)^2+\lambda_2(\phi_2^\dag\phi_2-V_2^2)^2\cr
&&+\lambda_3(\phi_1^\dag\phi_1-V_1^2+\phi_2^\dag\phi_2-V_2^2)^2+\lambda_4\left[(\phi_1^\dag\phi_1)(\phi_2^\dag\phi_2)-
(\phi_1^\dag\phi_2)(\phi_2^\dag\phi_1)\right]\cr
&&+(a\phi_1^\dag\phi_1+b\phi_2^\dag\phi_2)\phi^*\phi-c(\phi_1^\dag\phi_2\phi^2+\phi_2^\dag\phi_1\phi^{*2})
\ea
The $c$ term imposes the condition $-X_1+X_2+2X_\phi=0$. If we impose that the PQ current does 
not couple to the Goldstone boson that is eaten by the $Z$, 
we also get $X_1\cos^2\beta+X_2\sin^2\beta=0$. If furthermore we choose\footnote{There is arbitrariness
in this choice. This election conforms to the conventions existing in the literature.} $X_\phi=-1/2$ the 
PQ charges of the doublets are
\be\label{Xvalues}
X_1=-\sin^2\beta,\quad X_2=\cos^2\beta.
\ee

Global symmetries are not very evident in the way fields are introduced above. 
To remedy this let us define the matrices~\cite{ce}
\be
\Phi_{12}=(\tilde\phi_1~\phi_2)=\left(\begin{array}{cc}
\alpha_0^* & \beta_+\\
-\alpha_- & \beta_0
                                \end{array}\right),\quad
\Phi_{21}=(\tilde\phi_2~\phi_1)=\left(\begin{array}{cc}
\beta_0^* & \alpha_+\\
-\beta_- & \alpha_0
                                \end{array}\right)=\tau_2 \Phi_{12}^* \tau_2
\ee
and
\be
I=\Phi_{12}^\dag\Phi_{12}=\left(\begin{array}{cc}
 \phi_1^\dag\phi_1 & \tilde\phi_1^\dag\phi_2\\
 -\phi_1^\dag\tilde\phi_2 & \phi_2^\dag\phi_2
 \end{array}
 \right),\quad
J=\Phi_{12}^\dag\Phi_{21}=\phi_2^\dag\phi_1\mathbf{I}.
\label{eq:IandJ}
\ee
Defining also the constant matrix
$W=(V_1^2+V_2^2)\mathbf{I}/2+(V_1^2-V_2^2)\tau_3/2$, we can write the potential 
as
\ba
V(\phi,I,J)&=&\lambda_\phi(\phi^*\phi-V_\phi^2)^2+\frac{\lambda_1}4\left\{{\rm Tr}\left[(I-W)(1+\tau_3)\right]\right\}^2\cr
&&+\frac{\lambda_2}4\left\{{\rm Tr}\left[(I-W)(1-\tau_3)\right]\right\}^2+\lambda_3\left[{\rm Tr}(I-W)\right]^2\cr
&&+\frac{\lambda_4}4 {\rm Tr}\left[I^2-(I\tau_3)^2\right]+
\frac12 {\rm Tr}\left[(a+b)I+(a-b)I\tau_3\right]\phi^*\phi\cr
&&-\frac c2 {\rm Tr}(J\phi^2+J^\dag\phi^{*2}).
\label{eq:pot}
\ea

A $SU(2)_L\times SU(2)_R$ global transformation acts on our fields as
\be
\Phi_{ij}\to L\Phi_{ij}R^\dag,\quad I\to RIR^\dag,\quad J\to J.
\ee
We now we are in a better position to discuss the global symmetries of the
potential. The behavior of the different parameters under $SU(2)_R$ is shown in 
Table I. See also~\cite{wudka}.

\begin{table}
\begin{tabular}{|c|c|}
\hline
\textbf{Parameter} & \textbf{Custodial limit}\\
\hline
$\lambda_1,\,\lambda_2,\,\lambda_4$ & $\lambda_1=\lambda_2=\lambda$ and 
$\lambda_4=2\lambda$  \\
$\lambda_3$ & $\lambda_3$\\
$\lambda_\phi$ & $\lambda_\phi$\\
$V_1^2,\,V_2^2$  & $V_1^2=V_2^2=V^2$\\
$V_\phi$ & $V_\phi$\\
$a,\,b$ & $a=b$\\
$c$ & $c$\\
\hline
\end{tabular}
\caption{
In total, there are 11 parameters: 7 are custodially preserving and 
4 are custodially breaking. See the text for 
our usage of the expression `custodial symmetry' in the context of a 2HDM.\label{tab:cs}}
\end{table}

Finally, let us establish the action of the PQ symmetry previously discussed
 in this parametrization. Under the PQ transformation:
\be
\Phi_{12}\to\Phi_{12}e^{iX\theta},\quad\phi\to e^{iX_\phi\theta}\phi
\label{eq:phiPQ}
\ee
with
\be
X=\frac{X_2-X_1}{2}\mathbf{I}-\frac{X_2+X_1}{2}\tau_3,\quad
X_\phi=\frac{X_2-X_1}{2}
\ee
Using the values for $X_{1,2}$ given in Eq.~(\ref{Xvalues})
\be\label{pqcharge}
X=\left(\begin{array}{cc}
\sin^2\beta & 0\\
0 & \cos^2\beta
\end{array}\right),\quad X_{\phi}=-\frac12.
\ee

\section{Masses and mixings}\label{sec:mam}
We have two doublets and a singlet, so a total of $4+4+2=10$ spin-zero 
particles. Three particles are eaten by the $W^\pm$ and $Z$ 
and $7$ scalars fields are left in the spectrum; two charged Higgs, two $0^-$
states and three neutral $0^+$ states. 
Our field definitions will be worked out in full detail
in section~\ref{sec:nonlinear}. Here we want only to derive the spectrum. 
For the  charged Higgs  mass we have at tree level~\footnote{\label{note2}
Here and in the following we introduce the short-hand notation
$s^n_{m\beta} \equiv \sin^n(m\beta)$ and
$c^n_{m\beta} \equiv \cos^n(m\beta)$.}
\be\label{chargedmass}
m_{H_\pm}^2=8\left(\lambda_4v^2+\frac{cv_\phi^2}{s_{2\beta}}\right).
\ee

The quantity $v_\phi$ is proportional to the axion decay constant. Its value
is known to be very large (at least $10^7$ GeV and probably substantially 
larger $\sim 10^9$ GeV if all astrophysical constraints are taken into account,
see~\cite{axiondecay} for several experimental and cosmological bounds). 
It does definitely make
sense to organize the calculations as an expansion in $v/v_\phi$.

In the $0^-$ sector there are two degrees of freedom that mix with each 
other  with a mass matrix which has a vanishing eigenvalue. The eigenstate with zero mass is the axion   
and $A_0$  is the pseudoscalar Higgs with mass
\be\label{pseudoscalarmass}
m_{ A_0}^2=8c\left(\frac{v_\phi^2}{s_{2\beta}}+v^2 s_{2\beta}\right).
\ee
Eq.~(\ref{pseudoscalarmass}) implies $c\ge0$. 
For $c=0$, the mass matrix in the $0^-$ sector has a second
zero, i.e. in practice the $A_0$ field behaves as another axion.

In the $0^+$ sector, there are three neutral particles that mix with each other. 
With $h_i$  we denote the corresponding  $0^+$  mass eigenstates. The mass matrix is given in 
Appendix~\ref{app:B}. 
In the limit of large $v_\phi$, 
the mass matrix in the $0^+$ sector can be easily diagonalized~\cite{2hdmNatural} and presents one eigenvalue
 nominally of order $v^2$ and two of order $v_\phi^2$. Up to ${\cal O}(v^2/v_\phi^2)$, these masses are
\ba\label{mass1}
m_{h_1}^2&=&32v^2\left(\lambda_1c_\beta^4+\lambda_2s_\beta^4+\lambda_3\right)-16v^2\frac{\left(ac_\beta^2+bs_\beta^2-cs_{2\beta}\right)^2}{\lambda_\phi},\\
\label{mass11}
m_{h_2}^2&=&\frac{8c}{s_{2\beta}}v_\phi^2+8v^2s_{2\beta}^2(\lambda_1+\lambda_2)-4v^2\frac{\left[(a-b)s_{2\beta}+2cc_{2\beta}\right]^2}{\lambda_\phi-2c/s_{2\beta}},
\\
\label{mass111}
m_{h_3}^2&=&4\lambda_\phi v_\phi^2+4v^2\frac{\left[(a-b)s_{2\beta}+2cc_{2\beta}\right]^2}{\lambda_\phi-2c/s_{2\beta}}
+16v^2\frac{\left(ac_\beta^2+bs_\beta^2-cs_{2\beta}\right)^2}{\lambda_\phi}.
\ea
The field $h_1$ is naturally identified with the scalar boson of mass 125 GeV observed at the LHC.

It is worth it to stress that there are several situations where the above formulae are non-applicable, 
since the nominal expansion in powers of $v/v_\phi$ may fail. This may be the case where the coupling 
constants $a$, $b$, $c$ connecting
the singlet to the usual 2HDM are very small, of order say $v/v_\phi$ or $v^2/v_\phi^2$. One should also 
pay attention to the case $\lambda_\phi\to 0$ (we have termed this latter case as the `quasi-free singlet limit'). 
Leaving this last case aside, we have found that the above expressions for $m_{h_i}$ apply in the following situations:

\begin{itemize}
\item[] Case 1: The couplings $a$, $b$ and $c$ are generically of ${\cal O}(1)$,

\item[] Case 2: $a$, $b$ or $c$ are of ${\cal O}(v/v_\phi)$.

\item[] Case 3: $a$, $b$ or $c$ are of ${\cal O}(v^2/v_\phi^2)$ but $c \gg \lambda_i v^2/{v_\phi^2}$.
\end{itemize}

If $c \ll \lambda_i v^2/{v_\phi^2}$ the $0^-$ state is lighter than the lightest $0^+$ Higgs 
and this case is therefore already phenomenologically unacceptable. 
The only other case that deserves a 
separate discussion is 
\begin{itemize}
\item[] Case 4: Same as in case 3 but $c \sim \lambda_i {v^2}/{v_\phi^2}$ 
\end{itemize}
In this case, the masses in the $0^+$ sector read,  up to
${\cal O}(v^2/v^2_\phi)$, as
\be\label{mass2}
m^2_{h_1,h_2}=8v^2\left(K\mp\sqrt{K^2-L}\right)\:\text{and }\:
m^2_{h_3}=4\lambda_\phi v_\phi^2,
\ee
where 
\ba
K&=&2\left(\lambda_1c_\beta^2+\lambda_2s_\beta^2+\lambda_3\right)+\frac{c v_\phi^2}{2v^2 s_{2\beta}},\cr
L&=&4\left[\left(\lambda_1\lambda_2+\lambda_1\lambda_3+\lambda_2\lambda_3\right)s_{2\beta}^2+
\frac{c v_\phi^2}{v^2s_{2\beta}}\left(\lambda_1c_\beta^4+\lambda_2s_\beta^4+\lambda_3\right)\right].
\ea
Recall that here we assume $c$ to be of ${\cal O}(v^2/v^2_\phi)$. Note that
\be\label{sumrule}
m_{h_1}^2 +m_{h_2}^2 = 32
v^2\left(\lambda_1c_\beta^2+\lambda_2s_\beta^2+\lambda_3+\frac{c v_\phi^2}{4v^2
s_{2\beta}}\right).
\ee

In the `quasi-free singlet' limit, when $\lambda_\phi\to 0$ or more generically  $\lambda_\phi \ll a,b,c$
it is impossible to sustain the hierarchy $v\ll v_\phi$, so again this case is phenomenologically 
uninteresting (see Appendix~\ref{app:C} for details).

We note that once we set $\tan\beta$ to a fixed value, the
lightest Higgs to 125 GeV and $v_\phi$ to some large value
compatible with the experimental bounds, the mass spectrum in Eq.~(\ref{chargedmass}), (\ref{pseudoscalarmass}) and 
(\ref{mass1})-(\ref{mass111}) grossly depends on the parameters: $c$, $\lambda_4$ and $\lambda_\phi$, the latter
only affecting the third $0^+$ state that is anyway very heavy and definitely
out of reach of LHC experiments; therefore 
the spectrum depends on only two parameters. If case 4 is applicable, the situation is slightly
different and an additional combination of parameters dictates the mass of the second 
(lightish) $0^+$ state. This can be seen in the sum rule of Eq.~(\ref{sumrule}) after requiring that
$m_{h_1}= 125$ GeV. Actually this sum rule is also obeyed in cases 1, 2 and 3, but the
r.h.s is dominated then by the contribution from the parameter $c$ alone.

\subsection{Heavy and light states}
Here we want to discuss the spectrum of the theory according to the different scenarios that we have alluded to 
in the previous discussion.  Let us remember that it is always possible to identify one of the 
Higgses as the scalar boson found at LHC, namely $h_1$.
 
\begin{itemize}
\item[] Case 1. all Higgses except $h_1$ acquire a mass of order $v_\phi$. This includes the charged
and $0^-$ scalars, too. We term this situation `extreme decoupling'. The only light states are 
$h_1$, the gauge sector and the massless axion. This is the original DFSZ scenario
~\cite{dfs}

\item[] Case 2. This situation is similar to case 1 but now the typical scale of masses 
of $h_2$, $H_{\pm}$ and $A_0$
is
$\sqrt{vv_\phi}$. This range is beyond the LHC reach but it could perhaps be
explored with an accelerator in the 100 TeV region, a possibility being
currently debated. Again the only light particles are $h_1$, the axion and the
gauge sector. This possibility is natural in a technical sense as discussed
in~\cite{2hdmNatural} as an approximate extra symmetry would protect the
hierarchy.

\item[] Cases 3 and 4 are phenomenologically more interesting. Here we can at last have new states at the weak scale. 
In the $0^+$ sector, $h_3$ is definitely very heavy but $m^2_{h_1}$ and $m^2_{h_2}$ are
proportional to $v^2$ once we assume that  $c\sim v^2/v_\phi^2$. Depending on
the relative size of $\lambda_i $ and $c v^2_\phi/v^2$ one  would have to use
Eq.~(\ref{mass1}) or (\ref{mass2}). Because in case 3 one assumes that  $c
v^2_\phi/v^2$ is much larger than $\lambda_i$, $
h_1$ would still be the lightest
Higgs and $m_{h_2}$ could easily be in the TeV region.  When examining case 4 it
would be convenient to use the sum rule (\ref{sumrule}).  

We note that in case 4 the hierarchy between the different couplings is quite marked: typically to be 
realized one needs $c\sim 10^{-10} \lambda_i$, where $\lambda_i$ is a generic coupling of the potential.
It is the smallness of this number what results in the presence of light states at the weak scale. For a discussion
on the `naturalness' of this possibility see~\cite{2hdmNatural}.

 \end{itemize}

\subsection{Custodially symmetric potential}
While in the usual one doublet model, if  we neglect the Yukawa couplings and set the $U_Y(1)$ interactions to zero,
custodial symmetry is automatic, the latter is somewhat unnatural in 2HDM as one can write a fairly general potential. 
These terms are generically not invariant under global transformations $SU(2)_L\times 
SU(2)_R$ and therefore in the general case after the breaking  there is no custodial symmetry to speak of. 
Let us consider now the case where a global symmetry $SU(2)_L\times 
SU(2)_R$ is nevertheless present as there are probably good reasons to consider this limit. We may refer
somewhat improperly to this situation as to being `custodially symmetric' although after the breaking
custodial symmetry proper may or may not be present.
If $SU(2)_L\times SU(2)_R$ is to be a symmetry, 
the parameters of the potential have to be set according to the
custodial relations in Table~\ref{tab:cs}.
Now,  there are two possibilities  to spontaneously
break  $SU(2)_L\times SU(2)_R$ and to give mass to the gauge bosons.

\subsubsection{$SU(2)\times SU(2)\to U(1)$}
If the VEVs of the two Higgs fields are different ($\tan\beta\neq 1$), the custodial symmetry is 
spontaneously broken to $U(1)$. In this case, one can use the minimization equations of 
Appendix~\ref{sec:min} to eliminate 
$V$, $V_\phi$ and $c$  of Eq.~(\ref{eq:pot}). 
$c$ turns out to be of order $(v/v_\phi)^2$. In this case there are two extra Goldstone 
bosons: the charged Higgs is massless
\be
m_{H_\pm}^2=0.
\ee
Furthermore, the $A_0$ is light:
\be
m_{A_0}^2=16\lambda v^2\left(1+\frac{v^2}{v_\phi^2}s^2_{2\beta}\right)
\ee
This situation is clearly phenomenologically excluded.

\subsubsection{$SU(2)_L\times SU(2)_R\to SU(2)_V$}
In this case, the VEVs of the Higgs doublets are equal, so $\tan\beta=1$. The masses are
\be
m_{H_\pm}^2=8(2\lambda v^2+cv_\phi^2),\quad
m_{A_0}^2=8c(v^2+v_\phi^2)\quad\text{and}\quad m_{h_2}^2=m_{H_\pm}^2.
\ee
These three states are parametrically heavy, but they may be light in cases 3 and 4.\\ 
The rest of the $0^+$ mass matrix is $2\times2$ and has eigenvalues (up to second order in $v/v_\phi$)
\ba
m_{h_1}^2=16v^2\left[\lambda+2\lambda_3-\frac{(a-c)^2}{\lambda_\phi}\right]\quad\text{and}\quad
m_{h_3}^2=4\left[\lambda_\phi v_\phi^2+4v^2\frac{(a-c)^2}{\lambda_\phi}\right].
\ea

\begin{figure}[ht]
 \center
 \includegraphics[scale=0.5]{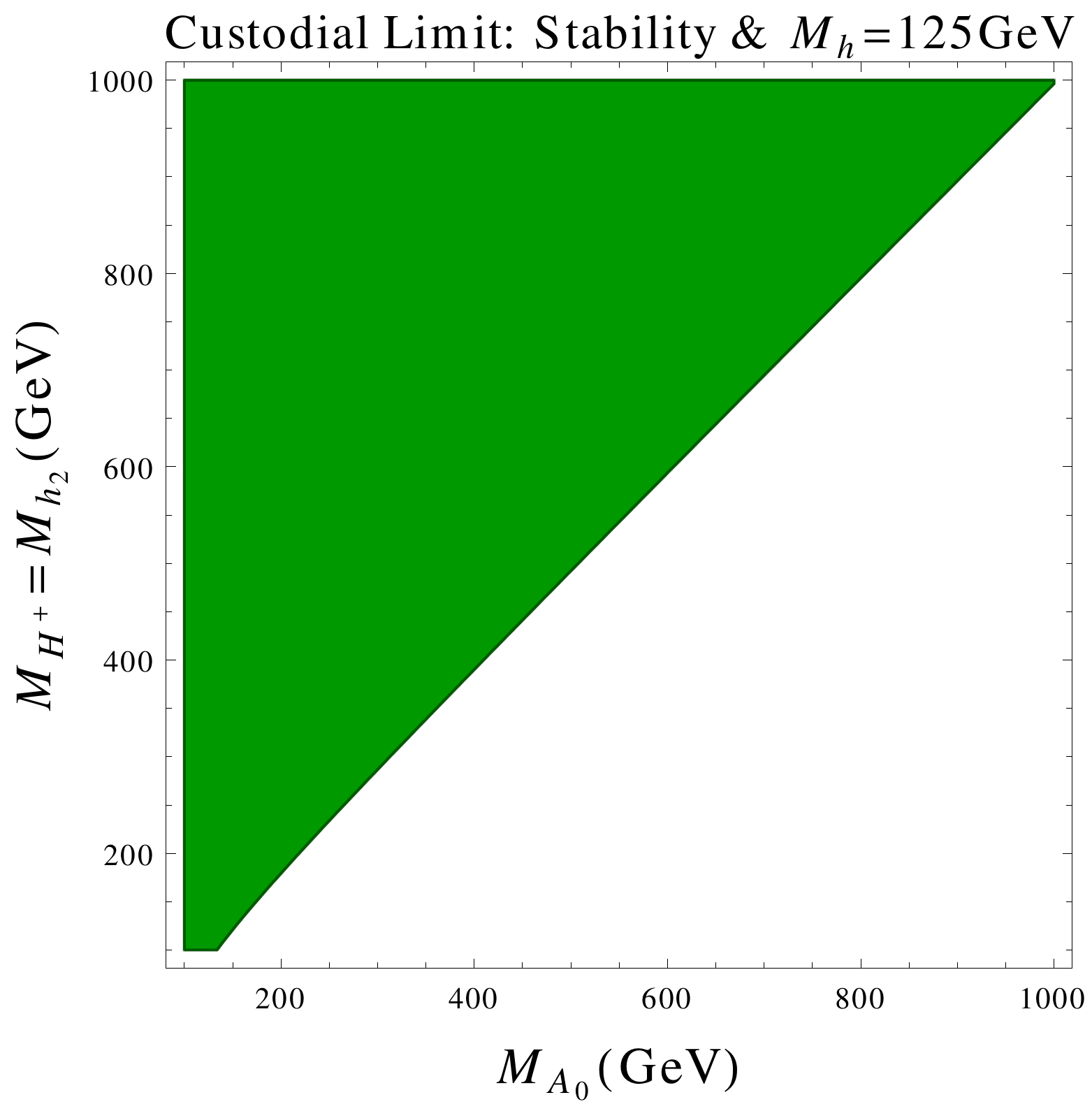}
 \caption{Dark/green: allowed region in the custodial limit after 
 requiring vacuum stability (see e.g. Appendix~\ref{sec:stab}). Each point
 in this region corresponds to a valid set of parameters in the DFSZ potential. Note that 
 $c$ is assumed to  be of order $ v^2/v_\phi^2$ and 
$c v^2_\phi/v^2$ has to be $\sim \lambda_i$ (case 4 discussed in the text).\label{fig:exclsu2}}
 \end{figure}

It is interesting to explore in this relatively simple case what sort of masses can be obtained
by varying the values of the couplings in the potential ($\lambda$, $\lambda_3$ and $c$). We
are basically interested in the possibility of obtaining a lightish spectrum (case 4 
previously discussed) and accordingly we assume that the natural scale of $c$ is $\sim v^2/v_\phi^2$.
We have to require the stability of the potential discussed in
Appendix~\ref{sec:stab} as well as $m_{h_1}=125$ GeV. The
allowed region is shown in Fig.~\ref{fig:exclsu2}.
Since we are in a custodially symmetric case there are no further restrictions to be obtained
from $\Delta\rho$.

\subsection{Understanding hierarchies}
As is well known in the MSM the Higgs mass has potentially large corrections
if the MSM is understood as an effective theory and one assumes that a larger scale
must show up in the theory at some moment. This is the case, for instance, if neutrino
masses are included via the see-saw mechanism, to name just a possibility. In this case
to keep the 125 GeV Higgs light one must do some amount of fine tuning.

In the DFSZ model such a large scale is indeed present from the outset and consequently
one has to envisage the possibility that the mass formulae previously derived may be
subject to large corrections due to the fact that $v_\phi$ leaks in the low energy
scalar spectrum. Let us discuss the relevance of the hierarchy problem in the different
cases discussed in this section.

In case 1 all masses in the scalar sector but the physical Higgs are heavy, of order $v_\phi$,
and due to the fact that the couplings $\lambda_i$ in the potential are generic (and
also the couplings $a,b,c$ connecting the two doublets to the singlet) the hierarchy may affect
the light Higgs quite severely and fine tuning of the $\lambda_i$ will be called for. However
this fine tuning is not essentially different from the one commonly 
accepted to be necessary in the MSM to keep the Higgs light if a new scale is somehow present.

In cases 3 and 4 the amount of additional fine tuning needed is none or very little. In these scenarios 
(particularly in case 4) the scalar spectrum is light, in the TeV region, and the only heavy
degree of freedom is contained in the modulus of the singlet. After diagonalization this results in a very heavy
$0^+$ state ($h_3$), with a mass or order $v_\phi$. However inspection of the potential reveals
that this degree of freedom couples to the physical sector with an strength $v^2/v_\phi^2$ and 
therefore may change the tree-level masses by a contribution of order $v$ ---perfectly acceptable.
In this sense the `natural' scenario proposed in \cite{2hdmNatural} does not apparently lead to
a severe hierarchy problem in spite of the large scale involved.

Case 2 is particularly interesting. In this case the intermediate masses are of order 
$\sqrt{v v_\phi}$, i.e. $\sim 100 $ TeV. There is still a very heavy mass eigenstate ($h_3$) but
again is nearly decoupled from the lightest Higgs as in cases 3 and 4. On the contrary, the
states with masses $\sim \sqrt{v v_\phi}$ do couple to the light Higgs with strength $\sim \lambda_i$ and
thus require ---thanks to the loop suppression factor--- only a very moderate amount of fine
tuning as compared to case 1.

It is specially relevant in the context of the hierarchy problem to consider the custodial case
discussed in the previous section. In the custodial limit the $A_0$ mass is protected as it
is proportial to the extended symmetry breaking parameter $c$. In addition
$m_{h_2}= m_{H^\pm}$. Should one wish to keep a control on radiative corrections, doing the fine tuning
necessary to keep $h_1$ and $h_2$ light should suffice and in fact
the contamination from the heavy $h_3$ is limited as said above. Of course, to satisfy the present 
data we have to worry only about $h_1$.

\section{Non-linear effective Lagrangian}\label{sec:nonlinear}
We have seen in the previous section that the spectrum of scalars resulting from the potential of
the  DFSZ model is generically heavy. It is somewhat difficult to have all the scalar masses at the weak
scale, although the additional scalars can be made to have masses in weak scale region in case
4.  The only exceptions are the three Goldstone bosons, the $h_1$ Higgs and the axion.  It is therefore somehow
natural to use a non-linear realization to describe the low energy sector formed by gauge bosons
(and their associated Goldstone bosons), the lightest Higgs $0^+$ state  $h_1$,  and the axion.
Deriving this effective action is one of the motivations of this work.

To construct the effective action we have to identify the proper variables and in order to do so
we will follow the technique described in~\cite{ce}. In that paper the case of a generic 2HDM
where all scalar fields were very massive was considered. Now we have to modify the method to
allow for a light state (the $h_1$) and also to include the axion degree of freedom.

We decompose the matrix-valued $\Phi_{12}$ field introduced in Section \ref{sec:mas} in the 
following form
\be
\Phi_{12}= {\cal U}{\cal M}_{12}.
\label{eq:um12}
\ee
${\cal U}$ is a $2\times 2$ matrix that contains the three Goldstone bosons associated to the breaking of 
$SU(2)_L$ (or more precisely of $SU(2)_L \times U(1)_Y$ to $U(1)_{em}$). We denote by $G^i$ these
Goldstone bosons
\be
{\cal U}=\exp\left(i\frac{\vec G\cdot\vec\tau}{v} \right).
\ee
Note that the matrices $I$ and $J$ of Eq.~(\ref{eq:IandJ}) entering the DFSZ potential are actually independent of ${\cal U}$.
This is immediate to see in the case of $I$ while for $J$ one has to use the property $\tau_2 {\cal U}^* = {\cal U}\tau_2$ valid
for $SU(2)$ matrices. The effective potential then does depend
only on the degrees of freedom contained in ${\cal M}_{12}$ 
whereas the Goldstone bosons drop from the potential, 
since, under a global  $SU(2)_L\times SU(2)_R$ rotation, $\Phi_{12}$ and ${\cal U}$  
transform as 
\be
\Phi_{12} \to L \Phi_{12} R^\dagger\quad {\cal U} \to L {\cal U} R^\dagger
\Rightarrow {\cal M}_{12} \to R {\cal M}_{12} R^\dagger.
\ee
Obviously the same applies to the locally gauged subgroup.

Let us now discuss the potential and ${\cal M}_{12}$  further. Inspection of 
the potential shows that because of the term proportional to $c$ the phase of the 
singlet field $\phi$ does not drop automatically from the potential and thus it cannot be
immediately identified with the axion. In other words, the phase of the $\phi$
field mixes with the usual $0^-$ scalar from the 2HDM.
To deal with this let us find a suitable phase both 
in ${\cal M}_{12}$ and in $\phi$ that drops from the effective potential -- this 
will single out the massless state to be identified with the axion. 

We write ${\cal M}_{12}=M_{12}U_a$, where $U_a$ is a unitary matrix containing the axion. 
An immediate choice is to take the generator of $U_a$ to be the identity, which
obviously can remove the phase of the singlet in the term in the 
effective potential proportional to $c$ while leaving the other terms manifestly
invariant. This does not exhaust all freedom however as we can include in the
exponent of $U_a$ a term proportional to $\tau_3$. It can be seen immediately
that this would again drop from all the terms in the effective potential,
including the one proportional to $c$ when taking into account that $\phi$
is a singlet under the action of $\tau_3$ that of course is nothing
but the hypercharge generator.
We will use the remaining freedom just discussed to properly 
normalize the axion and $A_0$ fields in the kinetic terms to which we now turn.

The gauge invariant kinetic term will be
\be\label{kinetic}
\mathcal{L}_{\rm{kin}}=\frac12(\partial_\mu\phi)^*\partial^\mu\phi+\frac14\rm{Tr}\left[(D_\mu\Phi_{12}^\dag)D^\mu\Phi_{12}\right],
\ee
where the covariant derivative is defined by
\be\label{covariantder}
D_\mu\Phi_{12}=\partial_\mu\Phi_{12}-i\frac g2\vec W_\mu\cdot\vec\tau\Phi_{12}+i\frac{g'}2B_\mu\Phi_{12}\tau_3.
\ee

By defining  $U_a=\exp\left(2 i a_\phi\,X /\vf\right)$ with $X$ in Eq.~(\ref{pqcharge}),
all terms in the kinetic term are diagonal and exhibit the canonical normalization. 
Moreover the field $a_\phi$ disappears from the potential.
Note that the phase redefinition implied in $U_a$ exactly coincides with 
the realization of the PQ symmetry on $\Phi_{12}$ in Eq.~(\ref{eq:phiPQ})  as is to be expected (this
identifies uniquely the axion degree of freedom). 

Finally,  the non-linear parametrization  of $\Phi_{12}$ reads as 
\be
\Phi_{12}= {\cal U}{M}_{12} U_a,
\label{eq:phi12onL}
\ee
with \ba
\!\!\!\!M_{12}=\sqrt2\left(
\begin{array}{cc}
(v+H)c_\beta-(S-i\frac{v_\phi}\vf A_0)s_\beta & \sqrt2 H_+ c_\beta\\
\sqrt2 H_- s_\beta & (v+H)s_\beta+(S+i\frac{v_\phi}\vf A_0)c_\beta
\end{array}
\right)
\label{eq:m12field}
\ea
and 
\be
v+H=\frac{c_\beta}{\sqrt2}\Re[\alpha_0]+\frac{s_\beta}{\sqrt2}\Re[\beta_0],\:
S=-\frac{s_\beta}{\sqrt2}\Re[\alpha_0]+\frac{c_\beta}{\sqrt2}\Re[\beta_0], \:
H_\pm=\frac{c_\beta\beta_\pm-s_\beta\alpha_\pm}{2}, 
\label{eq:HS}
\ee
in terms of the fields in Eq.~(\ref{eq:fields}). The singlet field $\phi$ is non-linearly parametrized as 
\be
\phi=\left(v_\phi+\rho-i\frac {v s_{2\beta}}\vf
A_0\right)\exp\left(i\frac{a_\phi}\vf\right).
\label{eq:phifield}\ee
With the parametrizations above the kinetic term is diagonal in terms of the fields of $M_{12}$ and 
$\rho$.
Moreover, the potential is independent of the axion and Goldstone bosons. 
All the fields appearing in Eqs.~(\ref{eq:m12field}) and~(\ref{eq:phifield}) have vanishing VEVs. 

Let us stress that $H$, $S$ and $\rho$ are not mass eigenstates and their relations 
with the $h_i$ mass 
eigenstates are defined through
\be
 H=\sum_{i=1}^3 R_{Hi}h_i,\quad S=\sum_{i=1}^3  R_{Si}h_i,\quad 
 \rho=\sum_{i=1}^3  R_{\rho i}h_i.
\label{eq:roth123}
\ee
The rotation matrix $R$ as well as  the corresponding mass matrix are given in Appendix~\ref{app:B}. 
$H$ and $S$ are the so called interaction eigenstates. 
In particular, $H$ couples to the gauge fields in the same way that the  usual MSM Higgs does. 

\subsection{Integrating out the heavy Higgs fields}
In this section we want to integrate out the heavy scalars in $\Phi_{12}$ of Eq.~(\ref{eq:phi12onL}) 
in order to build a low-energy effective theory at the TeV scale with an axion 
and a light Higgs.

As a first step, let us imagine that {{\em all} the states in $\Phi_{12}$ are heavy; 
upon their integration we will recover the Effective Chiral Lagrangian~\cite{efcl}
\be
\mathcal{L}= \frac{v^{2}}{4} {\rm Tr}\, D_{\mu}{\cal U}^{\dagger}D^{\mu}{\cal U}  + \sum_{i=0}^{13} a_{i} \mathcal{O}_i\, ,
\ee
where the $\mathcal{O}_i$ is a set of local gauge invariant operators~\cite{ey}, and the
symbol $D_\mu$ represents the covariant derivative defined in (\ref{covariantder}). The corresponding 
effective couplings $a_i$ collect the 
low energy information (up to energies $E\simeq 4\pi v$) pertaining to the heavy states integrated out. 
In the unitarity gauge, the 
term $D_{\mu}{\cal U}^{\dagger}D^{\mu}{\cal U}$ would generate 
the gauge boson masses.

If a light Higgs ($h=h_1$) and  axion are present, they have to be included explicitly 
as dynamical states~\cite{heff}, and the 
corresponding effective
Lagrangian will be (gauge terms are omitted in the present discussion)
\ba\label{chiralwithhiggs}
\mathcal{L} & = & 
\frac{v^{2}}{4} \left( 1+2 g_1\frac{h}{v}+ g_2 \frac{h^2}{v^2}+ \ldots\right) 
{\rm Tr}\, 
{\cal D}_{\mu}{\cal U}^{\dagger}{\cal D}^{\mu}{\cal U} \nn\\
&&+\left(\frac{v_\phi^2}{v_\phi^2+v^2s_{2\beta}}\right)\partial_\mu a_\phi \partial^\mu a_\phi+\frac{1}{2} \partial_{\mu} h \partial^{\mu} h  - V(h)  \\
&&+ \sum_{i=0}^{13} a_{i}\left(\frac{h}{v}\right)\mathcal{O}_i +\mathcal{L_{\text{ren}}},
\nn\ea  
where~\footnote{Note that the axion kinetic term is not well normalized in this expression yet. 
Extra contributions to the axion kinetic term also come from the term in the first line of Eq.~(\ref{chiralwithhiggs}). 
Only once we include these extra contributions, the axion kinetic term gets well normalized. See also discussion below.}
\be
{\cal D}_\mu {\cal U} = D_\mu {\cal U}+ {\cal U}(\partial_\mu U_a)U^\dagger_a
\label{eq:dcalU}
\ee 
formally amounting to a redefinition of the `right' gauge field and
\be
V(h) = \frac{m^2_h}{2} h^{2} - d_{3} (\lambda v)  h^{3} -  d_{4}
\frac{\lambda}{4} h^{4},
\ee
\be
\mathcal{L_{\text{ren}}} = \frac{c_1}{v^4}\left(\partial_\mu h \partial^\mu
h\right)^2 
+ \frac{c_2}{v^2}\left(\partial_\mu h\partial^\mu h\right){\rm Tr}\,{\cal D}_{\nu}{\cal U}^{\dagger}{\cal D}^{\nu}{\cal U}
+ \frac{c_3}{v^2}\left(\partial_\mu h\partial^\nu h\right){\rm Tr}\,{\cal D}^{\mu}{\cal
U}^{\dagger}{\cal D}_{\nu}{\cal U}.
\ee
Here $h$ is the lightest $0^+$ mass eigenstate, with mass $125$ GeV  
but couplings in principle different from the ones of a MSM Higgs. 
 The terms in  $\mathcal{L_{\text{ren}}}$ are
required for renormalizability~\cite{dobadollanes} 
at the one-loop level and play no role in the discussion.

The couplings $a_i$ are now functions of $h/v$, $ a_{i}({h}/{v})$, which are assumed to have 
a regular expansion and contribute
to different effective vertices. Their constant parts  $a_{i}(0)$ are related to the electroweak
precision parameters (`oblique corrections'). 

Let us see how the previous Lagrangian~(\ref{chiralwithhiggs}) can be derived. 
First, we integrate out from $\Phi_{12}= {\cal U}{M}_{12}U_a$ all heavy degrees of freedom, such as
$H^\pm$ and $A_0$, whereas we retain $H$ and $S$ because they contain a $h_1$ component, namely
 \ba
 \Phi_{12}= {\cal U}U_a {\overline M}_{12},\quad{ \overline M}_{12} = 
 \sqrt2\left(\begin{array}{cc}
 (v+H)c_\beta-Ss_\beta  & 0\\
 0 & (v+H)s_\beta+Sc_\beta
 \end{array}
 \right),
 \label{eq:m12bar}
\ea
where $H$ and $S$ stand respectively for $R_{H1} h_1$ and $R_{S1} h_1$. 

When the derivatives of the kinetic term of Eq.~(\ref{kinetic}) act on ${\overline M}_{12}$, we get the 
contribution $\partial_{\mu} h \partial^{\mu} h$ in 
Eq.~(\ref{chiralwithhiggs}).
Since the unitarity matrices, ${\cal U}$ and $U_a$ drop from the  potential of Eq.~(\ref{eq:pot}) only 
$V(h)$ remains.

To derive the first line of Eq.~(\ref{chiralwithhiggs}), we can use Eqs.~(\ref{eq:dcalU}) and ~(\ref{eq:m12bar}) 
to work out from the kinetic term of Eq.~(\ref{kinetic})
the contribution  
\be
 {\rm Tr}\,({\cal D}_{\mu}{\cal U}  { \overline M}_{12})^{\dagger}{\cal D}^{\mu}{\cal U} { \overline M}_{12}=
 \frac{v^{2}}{4} \left( 1+2\frac{H}{v}+ \ldots\right) 
{\rm Tr}\, 
{\cal D}_{\mu}{\cal U}^{\dagger}{\cal D}^{\mu}{\cal U} + {\cal L}(a_\phi,h).
\label{mam}
\ee
Here we used that $ {\overline M}_{12}  {\overline M}_{12}^\dagger $
has a piece proportional to the identity matrix and another proportional to $\tau_3$
that cannot contribute to the 
coupling with the gauge bosons since $ {\rm Tr}  D_{\mu}{\cal U}^{\dagger} D^{\mu}{\cal U} \tau_3 $ vanishes identically.
The linear contribution  in $S$ is of this type thus decoupling from 
the gauge sector and as a result only terms linear in $H$ survive.
Using that $[U_a,{ \overline M}_{12}]= [U_a, \tau_3]=0$, the matrix $U_a$ cancels out in all traces and the 
only remains of the axion in the low energy action is the modification $D_\mu \to {\cal D}_\mu$. 
The resulting effective action is invariant under global transformations ${\cal U}\to L{\cal U}R^\dagger$ but
now $R$ is an $SU(2)$ matrix only if custodial symmetry is preserved (i.e. $\tan\beta=1$). Otherwise the
right global symmetry group is reduced to the $U(1)$ subgroup. It commutes with $U(1)_{PQ}$. 

We then reproduce (\ref{chiralwithhiggs}) with $g_1=1$. However, this is true for the field $H$ on the l.h.s. of Eq.~(\ref{mam}), not
$h=h_1$ and this will reflect in a reduction in the value of the $g_i$ when one considers the coupling
to the lightest Higgs only.

A coupling among the $S$ field, the axion and the neutral Goldstone or the neutral gauge boson survives in Eq.~(\ref{mam}). 
This will be discussed in Section \ref{sec:higgs}.
As for the axion kinetic term, it is reconstructed with the proper normalization 
from the first term in (\ref{kinetic}) together
with a contribution from the `connection' $(\partial_\mu U_a)U_a^\dagger$ in
$ {\rm Tr}\,{\cal D}_{\mu}{\cal U}{\cal D}^{\mu}{\cal U} $ (see Eq.~(\ref{eq:Lsax}) in next section).
There are terms involving two axions and the Higgs that are not very relevant phenomenologically at this point.
This completes the derivation of the $O(p^2)$ terms in
the effective Lagrangian.

To go beyond this tree level and to determine the low energy constants  $a_i(0)$ in
particular requires a one-loop integration of the heavy degrees of freedom and matching the Green's functions
of the fundamental and the effective theories.

See e.g. \cite{efcl,ey} for a classification of all possible operators appearing up to ${\cal O}(p^6)$ that
are generated in this process. The information on physics beyond the MSM is encoded in the
coefficients of the effective chiral Lagrangian operators. Without including the (lightest) Higgs 
field $h$ (i.e. retaining only the constant term in the functions $a_i(h/v)$)
and ignoring the axion, there are
only two independent ${\cal O}(p^2)$ operators
\be
\label{p2}
{\cal L}^2= \frac{v^2}{4} {\rm Tr}(D_\mu {\cal U} D^\mu{\cal U}^\dagger) +
a_0(0) \frac{v^2}{4} ({\rm Tr}(\tau^3 {\cal U}^\dagger D_\mu{\cal U}))^2
\ee
The first one is universal, its coefficient being fixed by the $W$ mass.  As
we just saw it is flawlessly reproduced in the 2HDM at tree level
after assuming that the additional degrees of freedom are heavy. Loop
corrections do not modify it if $v$ is the physical Fermi scale.
The other one is related to the $\rho$ parameter.
In addition there are a few ${\cal O}(p^4)$ operators with their
corresponding coefficients
\be
\label{p4}
{\cal L}^4=\frac12 a_1(0) g g^\prime {\rm Tr}({\cal U} B_{\mu \nu} {\cal U}^\dagger
W^{\mu\nu})
-\frac14a_8(0)g^2{\rm Tr}({\cal U}\tau^3{\cal U}^\dagger W_{\mu\nu}){\rm Tr}(
{\cal U}\tau^3{\cal U}^\dagger W^{\mu\nu})+ ...
\ee
In the above expression $W_{\mu\nu}$ and $B_{\mu \nu}$ are the
field strength tensors associated to the $SU(2)_L$ and $U(1)_Y$ gauge fields, respectively.
In this paper we shall only consider the self-energy, or oblique,
corrections, which are  dominant in the 2HDM
model just as they are in the MSM.

The oblique corrections are often parametrized in terms
of the parameters $\varepsilon_1$, $\varepsilon_2$ and
$\varepsilon_3$  introduced in  \cite{AB}.
In an effective theory such as the one described by the Lagrangian
(\ref{p2}) and (\ref{p4}) $\varepsilon_1$,$\varepsilon_2$
and $\varepsilon_3$ receive one loop (universal)
contributions from the leading ${\cal O}(p^2)$ term $v^2{\rm Tr}(D_\mu
{\cal U}D^\mu {\cal U}^\dagger)$ and tree level contributions from
the $a_i(0)$. Thus
\be
\varepsilon_1= 2 a_0(0)+\ldots \qquad  \varepsilon_2= -g^2a_8(0) +\ldots
\qquad \varepsilon_3= -g^2a_1(0)+\ldots
\ee
where the dots symbolize the one-loop
${\cal O}(p^2)$ contributions. The latter
are totally independent of the specific symmetry breaking sector.
See e.g. \cite{ce} for more details.

A systematic integration of the heavy degrees of freedom, including the
lightest Higgs as external legs, would provide the dependence of the
low-energy coefficient functions on $h/v$, i.e. the form of the
functions $a_i(h/v)$. However this is of no interest to us here.

\section{Higgs and axion effective couplings}\label{sec:higgs}
The coupling of $h_1$ can be worked out from the one of $H$, which is exactly as in the MSM, namely
\be
g_1^{SM} H W_\mu W^\mu =g_1^{SM} (R_{H1} h_1 + R_{H2} h_2 + R_{H3}h_3  )W_\mu W^\mu
\ee
where $R_{H1}=1- (v/v_\phi)^2 A_{13}^2/2$ and $g_1^{SM}\equiv 1$. With the expression of $A_{13}$ given in Appendix~\ref{app:B}, 
\be
g_1= g_1^{SM} \times\left( 1- \frac{2v^2}{v_\phi^2 
\lambda_\phi^2}\left(ac^2_\beta+bs^2_\beta c_{2\beta} -  c s_{2\beta}\right)^2\right).
\ee
It is clear that in cases 1 to 3 
the correction to the lightest Higgs couplings to the gauge bosons are 
very small, experimentally indistinguishable from the MSM case. In any case
the correction is negative and $g_1< g_1^{SM}$.

Case 4 falls in a different category. Let us remember that this case corresponds
to the situation where $c \sim \lambda_i {v^2}/{v_\phi^2}$. Then the corresponding
rotation matrix is effectively $2\times 2$, with an angle $\theta$ that is
given in Appendix~\ref{app:B}. Then 
\be
g_1= g_1^{SM} \cos \theta.
\ee
In the custodial limit, $\lambda_1=\lambda_2$ and $\tan\beta =1$, this angle vanishes
exactly and $g_1= g_1^{SM}$. Otherwise this angle could have any value. Note however
that when $c \gg \lambda_i {v^2}/{v_\phi^2}$ then $\theta\to 0$ and the value
$g_1\simeq g_1^{SM}$ is recovered. This is expected as when $c$ grows case 4 moves into case 3.
Experimentally, from the LHC results we know~\cite{lhcbounds} that 
$g_1=[0.67,1.25]$ at $95\%$ CL.

Let us now discuss the Higgs-photon-photon coupling in this type of models. Let
us first consider the contribution from gauge and scalar fields in the
loop.  The diagrams
contributing to the coupling between the lightest scalar state $h_1$
and photons are exactly the same ones as in a generic 2HDM, via a loop
of gauge bosons and one of charged Higgses. In the DFSZ case the 
only change with respect to a generic 2HDM could be a modification in the $h_1WW$ (or 
Higgs-Goldstone bosons coupling) or in the $h_1 H^+ H^+$
tree-level couplings. The former has already been discussed while
the triple coupling of the lightest Higgs to two charged Higgses 
gets modified in the DFSZ model to
\ba
\lambda_{h_1H_+H_-}&=& 8vR_{H1}\left[(\lambda_1+\lambda_2)s^2_{2\beta}+4\lambda_3+2\lambda_4\right]
+16vs_{2\beta} R_{S1}\left(\lambda_2 c^2_\beta-\lambda_1s^2_\beta\right)\cr
& & +8v_\phi R_{\rho1}\left(as^2_\beta+bc^2_\beta-cs_{2\beta}\right).
\ea
Note that the first line involves only constants that are
already present in a generic 2HDM, while the second one does involve
the couplings $a,b$ and $c$ characteristic of the DFSZ model.

The coupling of the lightest Higgs to the up and down quarks is obtained from the Yukawa terms in Eq.~\eqref{yuk}
\be
\mathcal{L}(u,d,h_1)=\sqrt2h_1\left[G_1\left(R_{H1}c_\beta-R_{S1}s_\beta\right)\bar u_Lu_R+G_2\left(R_{H1}s_\beta+R_{S1}c_\beta\right)\bar d_Ld_R+\text{h.c.}\right]
\ee

The corresponding entries of the rotation matrix in the $0^+$ sector can be found 
in Appendix~\ref{app:B}. 
In cases 1, 2  and 3 the relevant entries are $R_{H1}\sim 1$, $R_{S1}\sim v^2/v_\phi^2$ and
$R_{\rho 1}\sim v/v_\phi$, respectively. Therefore the second term in the first line is
always negligible but the piece in the second one can give a sizeable contribution
if $c$ is of ${\cal O}(1)$ (case 1). This case could therefore be excluded or confirmed from
a precise determination of this coupling. In cases 2 and 3 this effective
coupling aligns itself with a generic 2HDM but with large (typically $\sim 100$ TeV) or moderately large 
(few TeV) charged Higgs masses.

Case 4 is slightly different again. In this case $R_{H1}=\cos\theta$ and
$R_{S1}=\sin \theta$ but $R_{\rho 1}=0$. The situation is again similar to a generic
2HDM, now with masses that can be made relatively light, but with a mixing angle
that because of the presence of the $c$ in (\ref{tan2theta}) terms may differ slightly from the
2HDM. For a review of current experimental fits in 2HDM the interested reader can 
see~\cite{pich}.

In this section  we will also list the tree-level couplings of the axion to the
light fields, thus completing the derivation of the effective low energy
theory. The tree-level couplings are very few actually as the axion
does not appear in the potential, and they are necessarily 
derivative in the bosonic part. From the kinetic term we get
\be
{\cal L}(a_\phi,h)=\frac{2R_{S1}}{\vf}h_1\partial^\mu a_\phi\left(\partial_\mu G_0+m_ZZ_\mu\right) + \text{ terms with 2 axions},
\label{eq:Lsax}
\ee

From the Yukawa terms \eqref{yuk} we also get
\be
{\cal L}(a_\phi,q,\bar q)=\frac{2i}{\vf}a_\phi\left(m_u s^2_\beta\bar u\gamma_5u+m_d c^2_\beta\bar d\gamma_5d\right).
\ee

The loop-induced couplings between the axion and gauge bosons 
(such as the anomaly-induced coupling $a_\phi\tilde F F$, of extreme importance for
direct axion detection~\cite{axiondecay}) will not be discussed here as they are
amply reported in the literature. 

\section{Matching the DFSZ model to 2HDM}\label{sec:matching}

The most general 2HDM potential can be  
read~\footnote{We have relabelled $\lambda\to\Lambda$ to avoid confusion with 
the potential of the DFSZ model.} e.g. from from \cite{2HDM2,pich}
\begin{alignat}{5}
 V(\phi_1,\phi_2) 
&= m^2_{11}\,\phi_1^\dagger\phi_1
 + m^2_{22}\,\phi_2^\dagger\phi_2
 - \left[ m^2_{12}\,\phi_1^\dagger\phi_2 + \text{h.c.} \right] \notag \\
&+ \frac{\Lambda_1}{2} \, (\phi_1^\dagger\phi_1)^2
 + \frac{\Lambda_2}{2} \, (\phi_2^\dagger\phi_2)^2
 + \Lambda_3 \, (\phi_1^\dagger\phi_1)\,(\phi_2^\dagger\phi_2) 
 + \Lambda_4 \, |\phi_1^\dagger\,\phi_2|^2 \notag \\
&+ \left[ \frac{\Lambda_5}{2} \, (\phi_1^\dagger\phi_2)^2 
        + \Lambda_6 \, (\phi_1^\dagger\phi_1) \, (\phi_1^\dagger\phi_2)
        + \Lambda_7 \, (\phi_2^\dagger\phi_2)\,(\phi_1^\dagger\phi_2) + \text{h.c.} 
   \right] \;.
\label{eq:2hdmpotential}
\end{alignat}
This potential contains 4 complex and 6 real parameters (i.e. 14 real numbers). 
The most popular 2HDM is obtained by imposing a ${Z}_2$ symmetry that is softly broken;
namely $\Lambda_6=\Lambda_7=0$  and $m_{12}\ne0$. The ${Z}_2$ approximate 
invariance helps remove flavour changing neutral current at tree-level.
A special role is played by the term proportional to $m_{12}$. 
This term softly breaks ${Z}_2$ but is necessary to control the decoupling limit of 
the additional scalars in a 2HDM and to eventually reproduce the MSM with a light
Higgs.
 
In the DFSZ model discussed here $v_\phi$ is very large and
at low energies the additional singlet field $\phi$ reduces approximately
to $\phi \simeq v_\phi \exp(a_\phi/v_\phi)$. Indeed, from (\ref{eq:phifield}) 
we see that $\phi$ has a  $A_0$ component but it can be in practice neglected
for an invisible axion since this component is $\propto v/v_\phi$. In addition the radial
variable $\rho$ can be safely integrated out. 

Thus, the low-energy effective theory defined by the DFSZ model
is a particular type of 2HDM model with the non-trivial
benefit of solving the strong $CP$ problem thanks  to the appearance of an
invisible axion\footnote{Recall that mass generation due to the anomalous
coupling with gluons  has not been considered in this work}.  
Indeed DFSZ reduces at low energy to a 2HDM containing 9 parameters in practice (see below, 
note that $v_\phi$ is used as input)
instead of the 14 of the general 2HDM case. 

The constants $\Lambda_{6,7}$ are absent as in many $Z_2$
invariant 2HDM but also  $\Lambda_{5}=0$. All these terms are   
not invariant under the Peccei-Quinn symmetry. In addition 
the $m_{12}$ that sofly breaks $Z_2$ and is necessary 
to control the decoupling to the MSM 
is dynamically generated by the PQ spontaneously symmetry breaking. 
There is no $\mu=m_{12}$ problem here concerning the naturalness of having non-vanishing $\mu$.  

At the electroweak scale 
the DFSZ potential of eq.~(\ref{potential}) can be matched to the 2HDM 
terms of (\ref{eq:2hdmpotential}) by the substitutions
\ba
m^2_{11}&=&\left[-2\lambda_1V_1^2+2
\lambda_3(V_1^2+V_2^2)+av_\phi^2\right]/4\\
m^2_{22}&=&\left[-2\lambda_1V_2^2+2
\lambda_3(V_1^2+V_2^2)+bv_\phi^2\right]/4\\
m_{12}^2&=&cv_\phi^2/4\\
\Lambda_1&=&(\lambda_1+\lambda_3)/8,\quad
\Lambda_2=(\lambda_2+\lambda_3)/8\\
\Lambda_3&=&(2\lambda_3+\lambda_4)/16,\quad
\Lambda_4=-\lambda_4/16,\\
\Lambda_5&=&0,\quad\Lambda_6=0,\quad\Lambda_7=0
\ea

Combinations of parameters of the DFSZ potential can be determined from
the four masses $m_{h_1}$, $m_{h_2}$, $m_{A_0}$ and $m_{H^+}$ and the two
parameters $g_1$ (or $\theta$) and $\lambda_{h_1H_+H_-}$ that controls
the Higgs-$WW$  and (indirectly) the Higgs-$\gamma\gamma$ couplings, whose expression 
in terms of the parameters of the potential has been given. As we have seen for
generic couplings, all masses but the lightest Higgs decouple and the
effective couplings take their MSM values.
In the phenomenologically
more interesting cases (cases 3 and 4) two of the remaining constants ($a$, $b$)  
drop in practice from the low-energy predictions and the effective 2HDM corresponding to  
DFSZ depends only on 7 parameters.  
If in addition custodial symmetry is assumed to be exact or nearly exact, 
the relevant parameters are actually totally determined by measuring three masses
and the two couplings ($m_{h_2}$ turns out to be equal to $m_{H^+}$ if custodial invariance
holds). Therefore LHC
has the potential of fully determining all the relevant parameters of
the DFSZ model. 
 
Eventually the LHC and perhaps a LC will be hopefully able to assess  
the parameters of the 2HDM potential and their symmetries to check the DFSZ relations. 
Of course finding a pattern of couplings in concordance with the pattern 
predicted by the low energy limit of DFSZ model would not yet prove the latter to 
be the correct microscopic theory as this would require measuring the 
axion couplings, which are not present in a 2HDM. In any case,
it should be obvious that the effective theory of the DFSZ is 
significantly more restrictive than a general 2HDM.

We emphasize that the above discussion refers mostly to case 4 as discussed in 
this work and it partly
applies to case 3 too. Cases 1 and 2 are in practice indistinguishable from
the MSM up to energies that are substantially larger from the ones currently 
accessible, apart from the presence of the axion itself. 
As we have seen, the DFSZ in this case is quite predictive and it does not 
correspond to a generic 2HDM but to one where massive scalars are all decoupled 
with the exception of the 125 GeV Higgs.

\section{Constraints from electroweak parameters}\label{sec:deltarho}
For the purposes of setting bounds on the masses of the new scalars
in the 2HDM, $\varepsilon_1=\Delta\rho$ is the most effective one. For this reason
we will postpone the analysis of $\varepsilon_2$ and $\varepsilon_3$
to a future publication.  

$\varepsilon_1$ can be computed by~\cite{AB}
\be
\varepsilon_1\equiv\frac{\Pi_{WW}(0)}{M_W^2}-\frac{\Pi_{ZZ}(0)}{M_Z^2},
\label{eq:deltarho}
\ee
with the gauge boson vacuum polarization functions defined as 
\be
\Pi^{\mu\nu}_{VV}(q)=g^{\mu\nu}\Pi_{VV}(q^2)+q^\mu q^\nu~{\rm terms}\quad (V=W,Z).
\ee
We need to compute loops of the type of Fig.~\ref{vvxy}. 
\begin{figure}[ht]
\center
\includegraphics[scale=0.6]{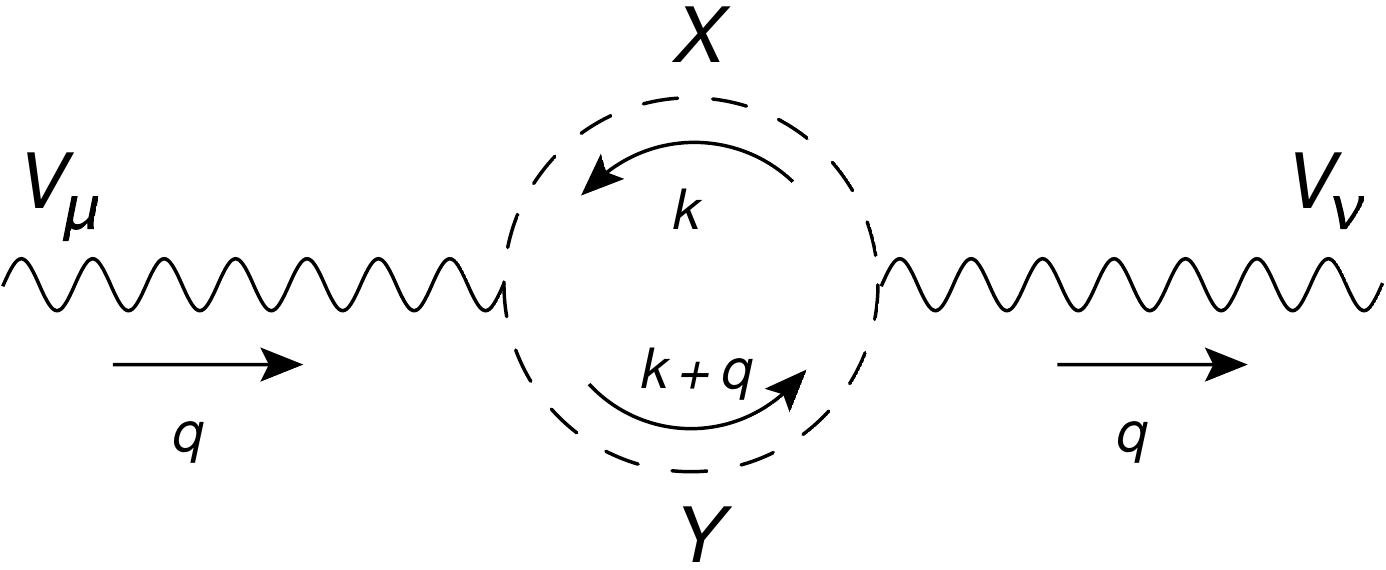}
\caption{Feynman diagram relevant for $\Pi^{\mu\nu}_{VV}(q)$.}\label{vvxy}
\end{figure}
These diagrams produce three kinds of terms. The terms proportional to two powers of 
the external momentum, $q_\mu q_\nu$, do not enter in $\Pi_{VV}(q^2)$. The terms
proportional to just one power vanish upon integration. Only the 
terms proportional to $k_\mu k_\nu$ survive and contribute.

Although it is an unessential approximation, to keep formulae relatively
simple we will compute $\varepsilon_1$ in the approximation $g^\prime=0$. The term
proportional to $(g^\prime)^2$ is 
actually the largest contribution in the MSM (leaving aside the 
breaking due to the Yukawa couplings) but it is only logarithmically
dependent on the masses of any putative scalar state and it can be safely omitted
for our purposes~\cite{ce}. The underlying reason is that in the 2HDM custodial
symmetry is `optional' in the scalar sector and it is natural to investigate
power-like contributions that would provide the strongest constraints. 
We obtain, in terms of the mass eigenstates and the rotation matrix of Eq.~(\ref{eq:roth123}),
\ba
\varepsilon_1&=& \frac{1}{16\pi^2 v^2}\Bigg[
m_{H_\pm}^2 - \frac{v^2_\phi}{\vvf} f(m_{H_\pm}^2,m_{A_0}^2) \\
&& + \sum_{i=1}^3   R^2_{Si} 
\left(\frac{v^2_\phi}{\vvf} f(m_{A_0}^2,m_{h_i}^2)-f(m_{H_\pm}^2,m_{h_i}^2)  \right)
\Bigg],
\ea
where $f(a,b)=a b/(b-a)\log b/a$ and $f(a,a)=a$. 
Setting $v_\phi\to \infty$ and keeping Higgs masses fixed, we formally recover the  $\Delta \rho$ 
expression in the 2HDM (see the Appendix in~\cite{ce}), namely
\ba
\varepsilon_1&=& \frac{1}{16\pi^2 v^2}\Bigg[
m_{H_\pm}^2 - f(m_{H_\pm}^2,m_{A_0}^2) + \sum_{i=1}^3   R^2_{Si} 
\left(f(m_{A_0}^2,m_{h_i}^2)-f(m_{H_\pm}^2,m_{h_i}^2)  \right)
\Bigg].
\ea
Now, in the limit $v_\phi\to \infty$ and  $m_{H_\pm}\to m_{A_0}$ (cases 1, 2 or 3 previously
discussed) the   $\Delta \rho$ above will go to zero as $v/v_\phi$ at least and the experimental bound is fulfilled automatically. 

However, we are particularly interested in case 4 that allows for a light spectrum
of new scalar states. We will study this in two steps. First we assume a 
`quasi-custodial' setting whereby we assume that 
custodial symmetry is broken {\em only}
via the coupling $\lambda_{4B}=\lambda_4-2\lambda$ being non-zero. 
Imposing vacuum stability (see e.g. Appendix~\ref{sec:stab}) and the experimental bound of 
$(\varepsilon_1-\varepsilon_1^{SM})/\alpha=\Delta T=0.08(7)$  
from the electroweak fits in~\cite{Dtexp} 
one gets the exclusion plots shown in Fig.~\ref{fig:ecl-su2}. 
\begin{figure}[ht]
\center
\includegraphics[scale=0.5]{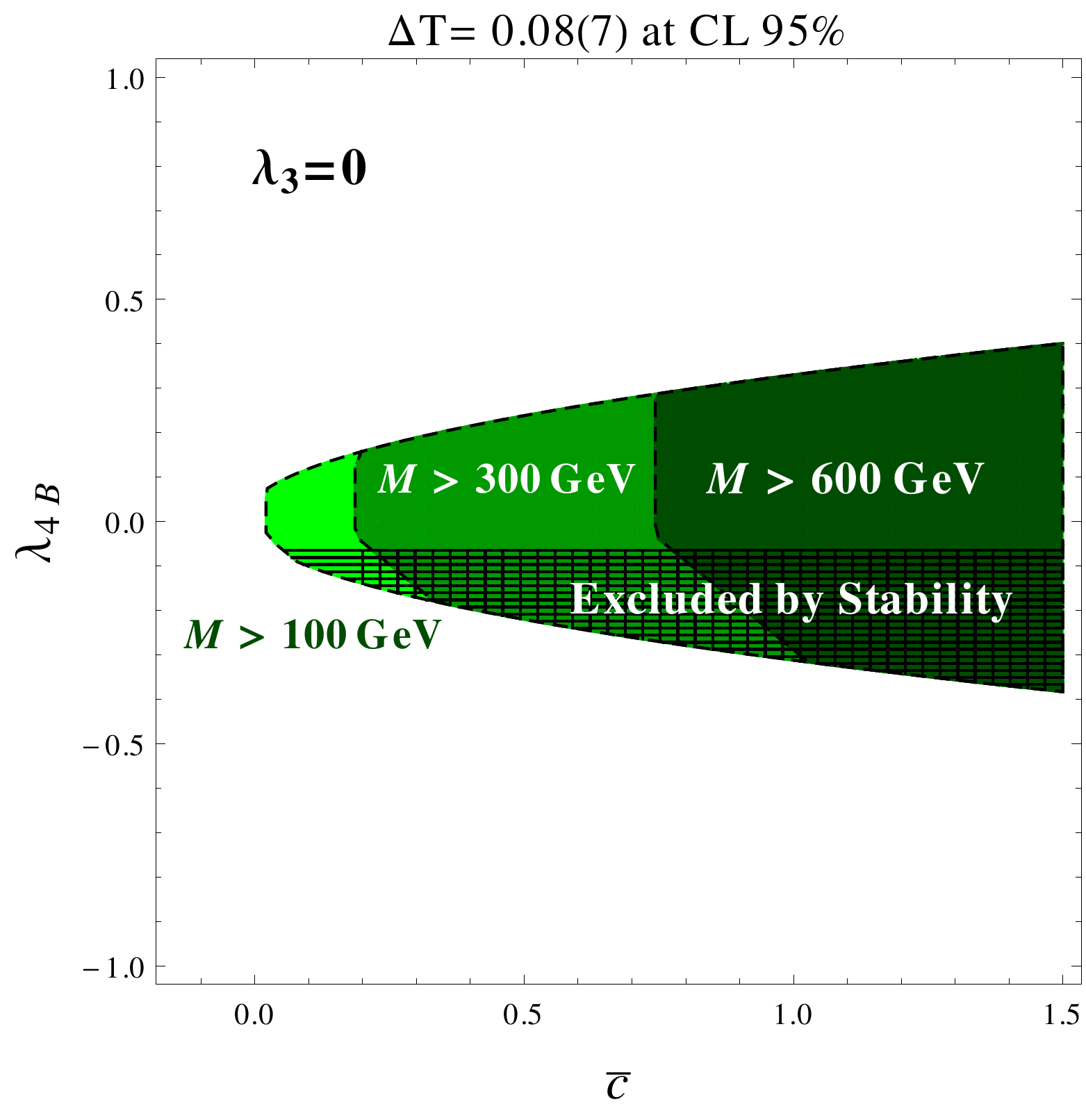}
\includegraphics[scale=0.5]{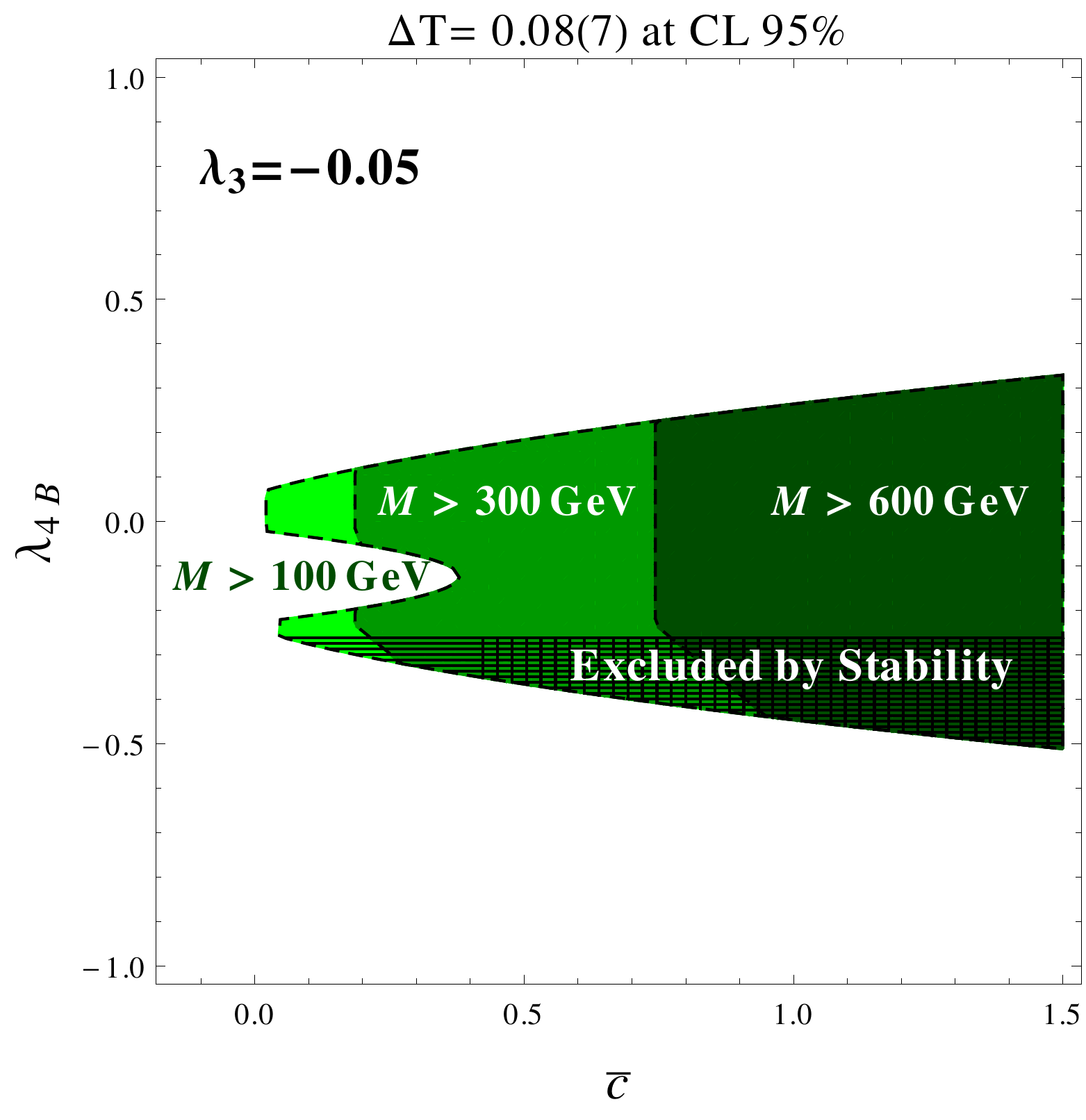}
\caption{Exclusion region for a `quasi-custodial' 2HDM potential 
with a $\lambda_{4B}=\lambda_4-2\lambda$ custodial breaking term. 
In this limit, the masses depend on the free parameters $\lambda_{4B}=\lambda_4-2\lambda$, $\lambda_{3}$ and 
$\bar c= c {v^2_\phi}/{v^2}$, and then the vacuum stability conditions of  Appendix~\ref{sec:stab} and $\Delta T$ 
can be used to exclude regions of the free parameters above. The left, right plots are for $\lambda_{3}=-0.05$, 
$\lambda_{3}=0$, respectively.
Different color regions imply different cuts assuming that all masses 
($m_{A_0}$, $m_{H_\pm}$ and $m_{h_2}$) are greater than 100, 300 or 600 GeV (light to dark).
The potential becomes unstable for $\lambda_3 > 0.03$. }
\label{fig:ecl-su2}
\end{figure}

It is also interesting to show (in this same `quasi-custodial' limit) the range
of masses allowed by the present constraints on $\Delta T$, without any reference to
the parameters in the potential. This is shown for two reference values of $m_{A_0}$ in Fig.~\ref{eclM-su2}. Note
the severe constraints due to the requirement of vacuum stability.  
\begin{figure}[ht]
\center
\includegraphics[scale=0.5]{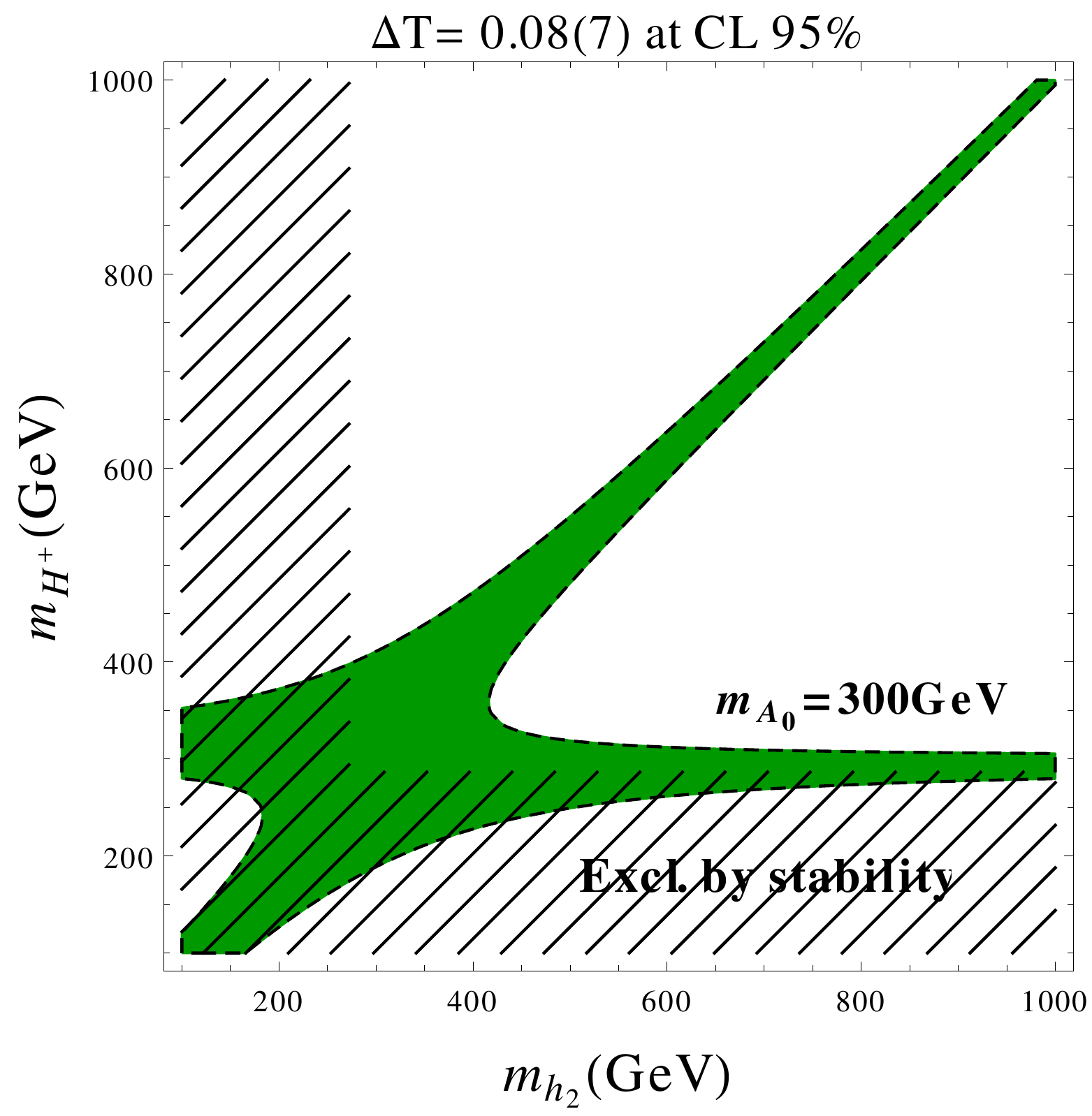}
\includegraphics[scale=0.5]{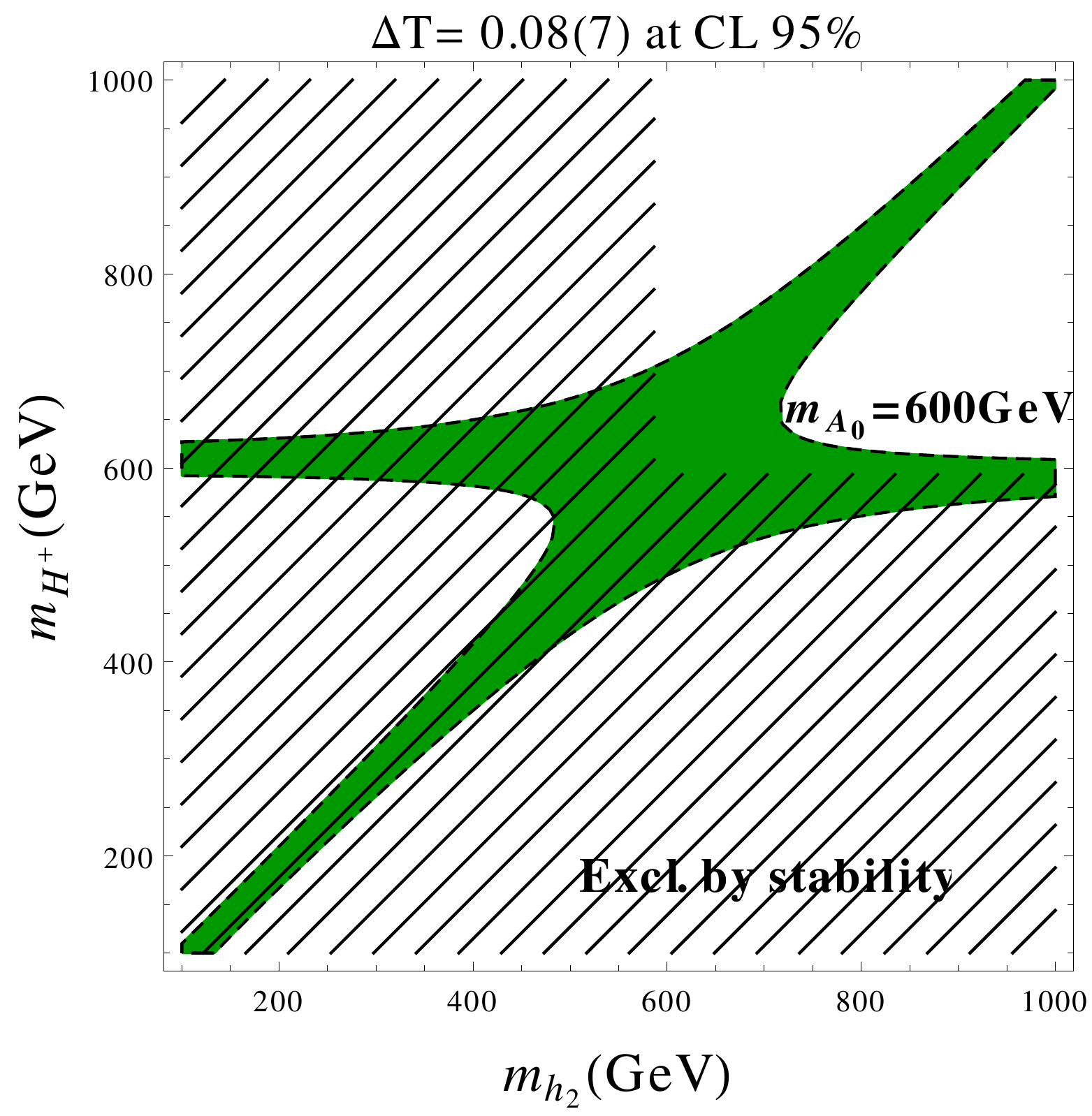}
\caption{Exclusion plot imposed by the constraint from $\Delta T$ on the second
$0^+$ state (i.e. `second Higgs') and the charged Higgs masses for two reference values of
$M_{A_0}$ (left: 300 GeV, right: 600 GeV)  in the `quasi-custodial' case discussed in Fig.~\ref{fig:ecl-su2}.
The concentration of points along approximately two axis 
is easy to understand 
after inspection of the relevant formula for $\Delta T$.  $\Delta T$ is vanishing at the custodial limit $M_{h_2}=M_{H^\pm}$, and also for 
$M_{H^\pm}=M_{A_0}$.  The regions excluded by considerations
of stability of the potential are shown.}
\label{eclM-su2}
\end{figure}

Finally let us turn to the consideration of the general case 4. We now completely give up 
custodial symmetry and hence the three masses $m_{A_0}$, $m_{H_\pm}$ and $m_{h_2}$ are unrelated, except for the eventual lack of stability of the potential. 
In this case, the rotation $R$ can be different form the identity which was the case in the `quasi-custodial' scenario above.
In particular, $R_{S2}=\cos\theta$ from Appendix~\ref{app:B} and the angle 
$\theta$ is not vanishing. However, experimentally 
$\cos \theta$ is known to be
very close to one (see section \ref{sec:higgs}). If we assume that
$\cos\theta$ is exactly equal to one, we get the exclusion/acceptance regions shown
in Fig. \ref{eclM-gen}. Finally, Fig. \ref{eclM-theta} depicts the analogous plot for
$\cos\theta=0.95$ that is still allowed by existing constraints. We wee that the allowed range
of masses are much more severely restricted in this case.

\begin{figure}[ht]
\center
\includegraphics[scale=0.5]{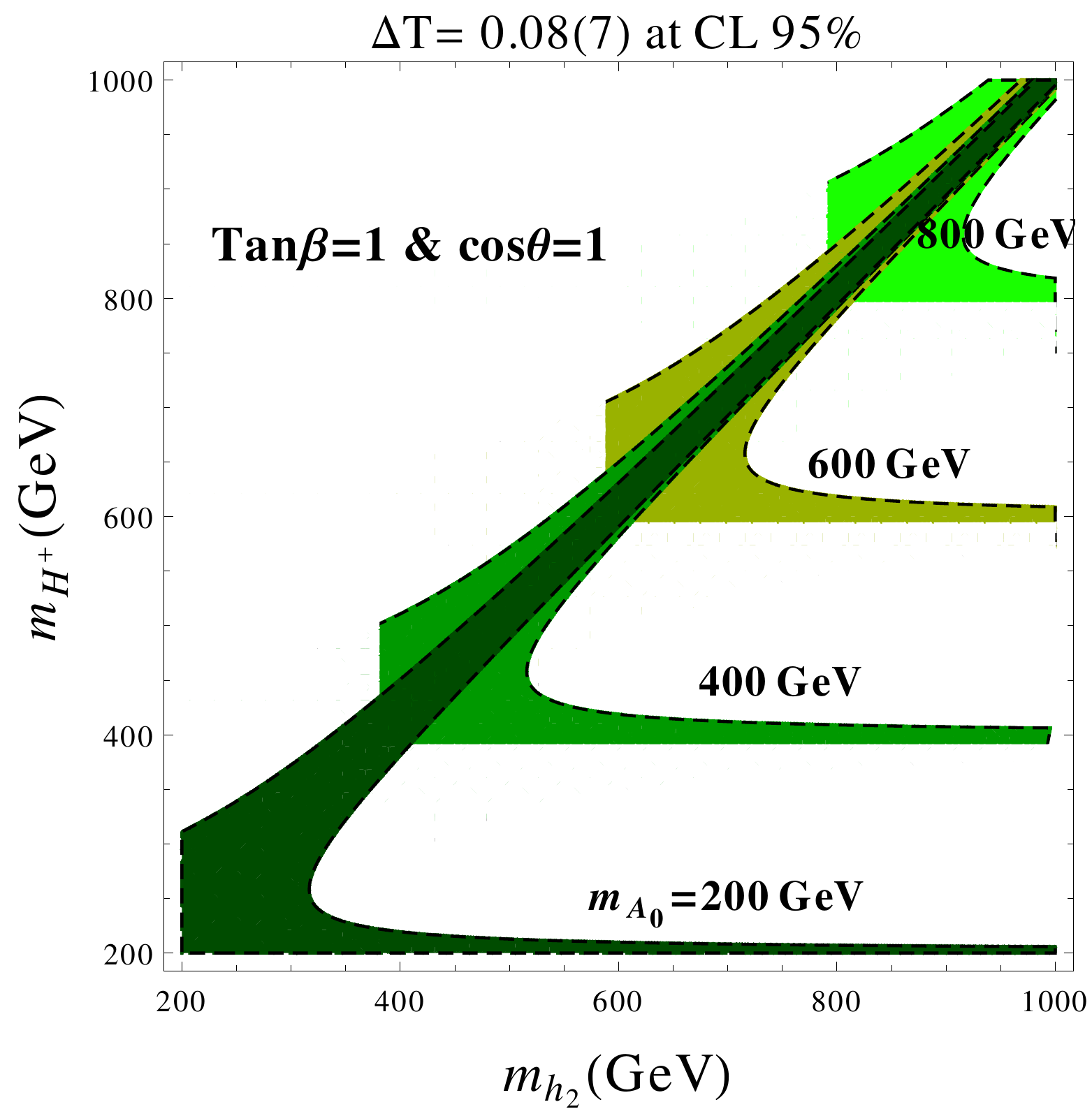}
\includegraphics[scale=0.5]{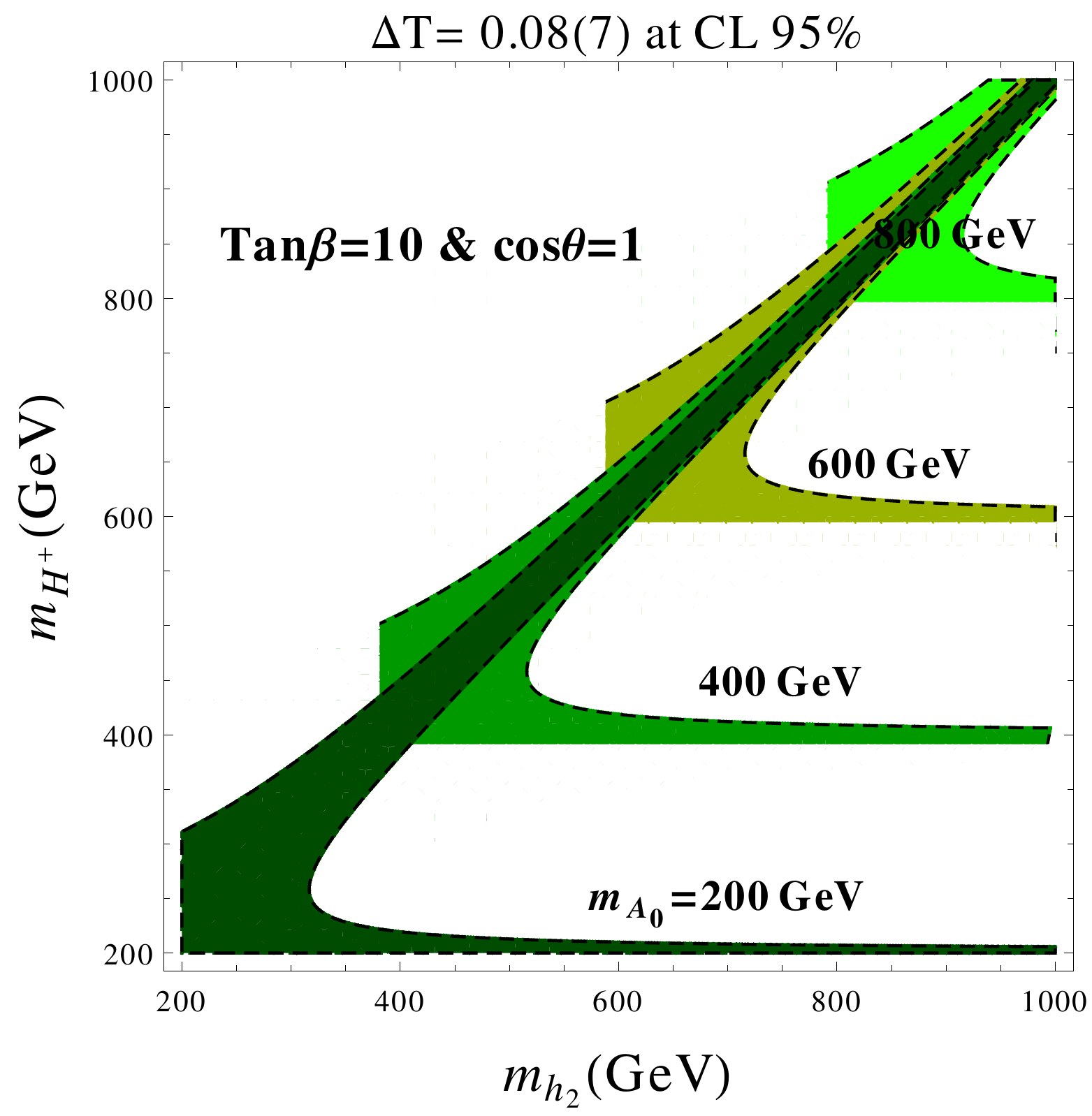}
\caption{Exclusion plot imposed by the constraint from $\Delta T$ on the second
$0^+$ state (i.e. `second Higgs') and the charged Higgs masses for several reference 
values of
$m_{A_0}$ and $\tan\beta$ in the general case. The value $\cos\theta=1$ is assumed here. The successive
horizontal bands correspond to different values of $m_{A_0}$. The stability bounds have already been implemented,  
effectively cutting off the left and lower arms of the regions otherwise acceptable. }
\label{eclM-gen}
\end{figure}

\begin{figure}[ht]
\center
\includegraphics[scale=0.5]{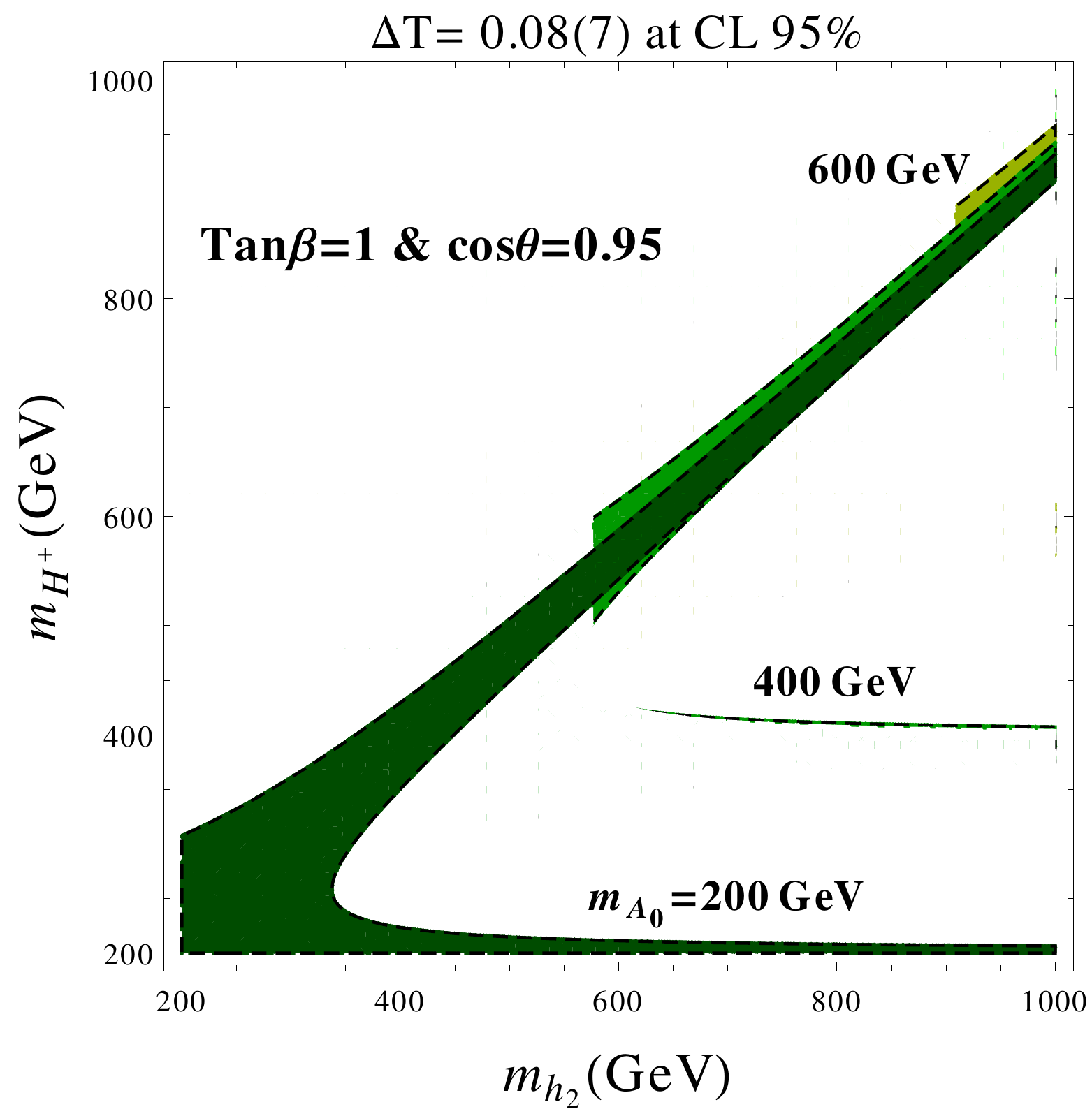}
\caption{Exclusion plot imposed by the constraint from $\Delta T$ on the second
$0^+$ state (i.e. `second Higgs') and the charged Higgs masses for several reference values of
$m_{A_0}$. Here we take $\tan\beta=1$ and allow $\cos\theta=0.95$, which is
consistent with present constraints. }
\label{eclM-theta}
\end{figure}

\section{Conclusions}

With the LHC experiments gathering more data, the exploration of the symmetry breaking sector of 
the Standard Model will gain renewed impetus. Likewise, it is important to search for dark matter
candidates as this is a degree of freedom certainly missing in the minimal Standard Model. An 
invisible axion is an interesting candidate for dark matter; however trying 
to look for direct evidence of its existence at the LHC is hopeless as it is extremely weakly coupled. 
Therefore we have to resort to less direct ways to explore this sector
by formulating consistent models that include the axion and deriving 
consequences that could be experimentally tested.

In this work we have explored such consequences in the DFSZ model, an extension of the popular 2HDM.
A necessary characteristic of models with an invisible axion is the presence of the Peccei-Quinn
symmetry. This restricts the form of the effective potential. We have taken into account the recent
data on the Higgs mass and several effective couplings, and included the constraints from electroweak
precision parameters.

Four possible scenarios have been considered. In virtually all points of
parameter space of the DFSZ model we do not really expect to see any relevant
modifications with respect to the  minimal Standard Model predictions. The new
scalars have masses of order $v_\phi$ or $\sqrt{v v_\phi}$ in two of the cases
discussed. The latter could perhaps be reachable with a $100$ TeV circular
collider although this is not totally guaranteed. In a third case, it would be
possible to get scalars in the  multi-TeV region, making this case testable in
the future at the LHC.  Finally, we have identified a fourth situation where a
relatively light spectrum emerges. The last two cases correspond to a situation
where the coupling  between the singlet and the two doublets is of order
$v^2/v_\phi^2$; i.e. very small ($10^{-10}$ or less) and in order to get a
relatively light spectrum in addition one has to 
require some couplings to be commensurate (but not necessarily fine-tuned). 

The fact that some specific couplings are required to be very small may seem odd, but as it has been
argued elsewhere it is technically natural, as the couplings in question do break some extended
symmetry and are therefore protected. From this point of view these values are perfectly acceptable.

The results on the scalar spectrum are derived here at tree level only and are of course subject to large radiative
corrections in principle. However one should note two ingredients that should ameliorate the hierarchy problem. The
first observation is that the mass of the $0^-$ scalar is directly proportional to $c$; it is exactly zero
if the additional symmetries discussed in \cite{2hdmNatural} hold. It is therefore somehow protected. On the
other hand custodial symmetry relates different masses, helping to maintain other relations. Some hierarchy
problem should still remain but of a magnitude very similar to the one already present in the minimal Standard
Model. 

We have imposed on the model known constraints such as the fulfilment of the bounds on the $\rho$-parameter. These
bounds turn out to be automatically fulfilled in most of parameter space and become only relevant when 
the spectrum is light (case 4). This is particularly relevant as custodial symmetry is by no means automatic
in the 2HDM. Somehow the introduction of the axion and the related Peccei-Quinn symmetry makes possible custodially
violating consequences naturally small. We have also considered the experimental bounds on the Higgs-gauge bosons and
Higgs-two photons couplings. Together with four scalar masses, these parameters determine in an almost unique way
the DFSZ potential, thus showing that it has subtantial less room to maneuver than a generic 2HDM. 

In conclusion, DFSZ models containing an invisible axion are natural and, in spite of the large
scale that appears in the model to make the axion nearly invisible, there is the possibility that
they lead to a spectrum that can be tested at the LHC. This spectrum is severely constrained, 
making it easier to prove or disprove such possibility in the near future. 
On the other hand it is perhaps more likely that the new states predicted by the model
lie beyond the LHC range. In this situation the model hides itself by making 
indirect contributions to most observables quite small.

 \section*{Acknowledgements}
This work is supported by grants FPA2013-46570, 2014-SGR-104 and Consolider grant CSD2007-00042 (CPAN). 
A. Renau acknowledges the
financial support of a FPU pre-doctoral grant.

\appendix
\section{Minimization conditions of the potential}
\label{sec:min}
The minimization conditions for the potential \eqref{potential} are
\be
\lambda_1\left(2v^2c_\beta^2-V_1^2\right)+\lambda_3\left(2v^2-V_1^2-V_2^2\right)+\frac{v_\phi^2}2\left(a+c\tan\beta\right)=0,
\ee
\be
\lambda_2\left(2v^2s_\beta^2-V_2^2\right)+\lambda_3\left(2v^2-V_1^2-V_2^2\right)+\frac{v_\phi^2}2\left(b+\frac{c}{\tan\beta}\right)=0,
\ee
\be
\lambda_\phi\left(v_\phi^2-V_\phi^2\right)+2v^2\left(ac_\beta^2+bs_\beta^2-cs_{2\beta}\right)=0.
\ee
These allow us to eliminate the dimensionful parameters
$V_\phi$, $V_1$ and $V_2$ in favor of the different couplings, $v$ and $v_\phi$. In the case
where $\lambda_\phi=0$ it is also possible to eliminate $c$.

\section{$0^+$ neutral sector mass matrix}
\label{app:B}
The $3\times 3$ mass matrix is
\be
M_{HS\rho}=4
\left(
\begin{array}{ccc}
8v^2\left(\lambda_1c_\beta^4+\lambda_2s_\beta^4+\lambda_3\right) & 4v^2\left(-\lambda_1c_\beta^2+\lambda_2s_\beta^2\right)s_{2\beta} 
& 2vv_\phi\left(ac_\beta^2+bs_\beta^2-cs_{2\beta}\right)\\
 4v^2\left(-\lambda_1c_\beta^2+\lambda_2s_\beta^2\right)s_{2\beta}  & \frac{2cv_\phi^2}{s_{2\beta}}+2v^2\left(\lambda_1+\lambda_2\right)s_{2\beta}^2 
 & -vv_\phi\left[\left(a-b\right)s_{2\beta}+2cc_{2\beta}\right] \\
2vv_\phi\left(ac_\beta^2+bs_\beta^2-cs_{2\beta}\right) & -vv_\phi\left[\left(a-b\right)s_{2\beta}+2cc_{2\beta}\right] & \lambda_\phi v_\phi^2
\end{array}
\right)
\ee
This is diagonalized with a rotation
\be
\left(\begin{array}{c}
H\\
S\\
\rho
\end{array}\right)=R
\left(\begin{array}{c}
h_1\\h_2\\h_3
\end{array}\right).
\ee
We write the rotation matrix as
\be
R=\exp\left(\frac v{v_\phi}A+\frac{v^2}{v_\phi^2}B\right),\quad A^T=-A,\quad B^T=-B
\ee
and work up to second order in ${v}/{v_\phi}$. We find
\be
A_{12}=B_{13}=B_{23}=0,
\ee
so the matrix is
\be
R=\left(
\begin{array}{ccc}
1-\frac{v^2}{v_\phi^2}\frac{A_{13}^2}{2} & -\frac{v^2}{v_\phi^2}\frac{A_{13}A_{23}-2B_{12}}{2} & \frac{v}{v_\phi}A_{13}\\
-\frac{v^2}{v_\phi^2}\frac{A_{13}A_{23}+2B_{12}}{2} & 1-\frac{v^2}{v_\phi^2}\frac{A_{23}^2}{2} & \frac{v}{v_\phi}A_{23}\\
-\frac{v}{v_\phi}A_{13} & -\frac{v}{v_\phi}A_{23}  &  1-\frac{v^2}{v_\phi^2}\frac{A_{13}^2+A_{23}^2}{2}
\end{array}
\right),
\ee
with
\be
A_{13}=\frac2{\lambda_\phi}\left(ac_\beta^2+bs_\beta^2-cs_{2\beta}\right),\:
A_{23}=\frac{(a-b)s_{2\beta}+2cc_{2\beta}}{\frac{2c}{s_{2\beta}}-\lambda_\phi},
\ee
\be
B_{12}=
-\frac2cs^2_{2\beta}\left(\lambda_1c_\beta^2-\lambda_2s_\beta^2\right)
+\frac{s_{2\beta}}{\lambda_\phi c}\frac{c-\lambda_\phi s_{2\beta}}{2c-\lambda_\phi s_{2\beta}}
\left(ac_\beta^2+bs_\beta^2-cs_{2\beta}\right)\left[(a-b)s_{2\beta}+2cc_{2\beta}\right]
\ee
In the case of section \ref{sec:mam} when the breaking of 
custodial symmetry is $SU(2)\times SU(2)\to U(1)$ the mass matrix is
\be
M_{HS\rho}=4\left(\begin{array}{ccc}
     8v^2\left[\lambda_3+\lambda(s^4_\beta+c^4_\beta)\right] & -2\lambda v^2 s_{4\beta} & 2vv_\phi\left(a+2\lambda\frac{v^2}{v_\phi^2} s^2_{2\beta}\right)  \\
     -2\lambda v^2 s_{4\beta} & -4\lambda v^2 c^2_{2\beta} & 2\lambda\frac{v^3}{v_\phi} s_{4\beta} \\
     2vv_\phi\left(a+2\lambda\frac{v^2}{v_\phi^2} s^2_{2\beta}\right) & 2\lambda\frac{v^3}{v_\phi} s_{4\beta} & \lambda v_\phi^2
   \end{array}\right).
\ee
For case 4 of section \ref{sec:mam} the rotation matrix is
\be\label{tan2theta}
R=\left(
\begin{array}{ccc}
\cos\theta & -\sin\theta & 0\\
\sin\theta & \cos\theta & 0\\
0 & 0 & 1
\end{array}
\right),\quad
\tan2\theta=
-\frac{\left(\lambda_1c_\beta^2-\lambda_2s_\beta^2\right)s_{2\beta}}{\left(\lambda_1c_\beta^2-\lambda_2s_\beta^2\right)c_{2\beta}+\lambda_3-
{cv_\phi^2}/(4v^2s_{2\beta})}.
\ee

\section{The limit $\lambda_\phi=0$}
\label{app:C}
The eigenvalues of  the $3\times 3$ mass matrix in the $0^+$ sector are
\ba
m_{h_1}^2&=&32v^2\left(\lambda_1c^4_\beta+\lambda_2s^4_\beta+\lambda_3\right)\cr
m_{h_2}^2&=&\frac{v_\phi^2}2s^2_{2\beta}(ac^2_\beta+bs^2_\beta)+\mathcal{O}(v^2)\cr
m_{h_3}^2&=&-8v^2\frac{(ac_\beta-bs_\beta)^2}{ac^2_\beta+bs^2_\beta}
\ea
Either $m_2^2$ or $m_3^2$ is negative. Note that the limit of $a,b$ small can not be taken directly in this case.

\section{Vacuum stability conditions and mass relations}
\label{sec:stab}
Vacuum stability implies the following conditions  on the
parameters of the potential~\cite{2HDM2}:
\be\nonumber
\lambda_1+\lambda_3>0,\quad \lambda_2+\lambda_3>0,\quad
2\lambda_3+\lambda_4+2\sqrt{(\lambda_1+\lambda_3)(\lambda_2+\lambda_3)}>0,
\ee
\be 
\lambda_3+\sqrt{(\lambda_1+\lambda_3)(\lambda_2+\lambda_3)}>0.
\ee
In the case of custodial symmetry except for $\lambda_{4B}\ne 0$, 
these conditions reduce to
\be
\lambda+\lambda_3>0,\quad \lambda+2\lambda_3>0,\quad 
4\lambda+4\lambda_3+\lambda_{4B}>0
\ee
and assuming $a,\,b,\,c$ very small (e.g. case 4) they impose two conditions on the masses for :
\be
m_{A_0}^2+m_{h_1}^2-m_{h_2}^2>0,\quad m_{H_\pm}^2+m_{h_1}^2-m_{A_0}^2>0.
\ee

\section{Vertices and Feynman Rules in the DFSZ model}
In the limit of $g'=0$, all the diagrams involved in the calculation of $\varepsilon_1$ are of the type of Fig.~\ref{vvxy}. 
All the relevant vertices are of the type seen in Fig.~\ref{vxy}, with all momenta assumed to be incoming. 
The relevant Feynman rules are as follows:
\begin{figure}[ht]
\center
\includegraphics[scale=0.6]{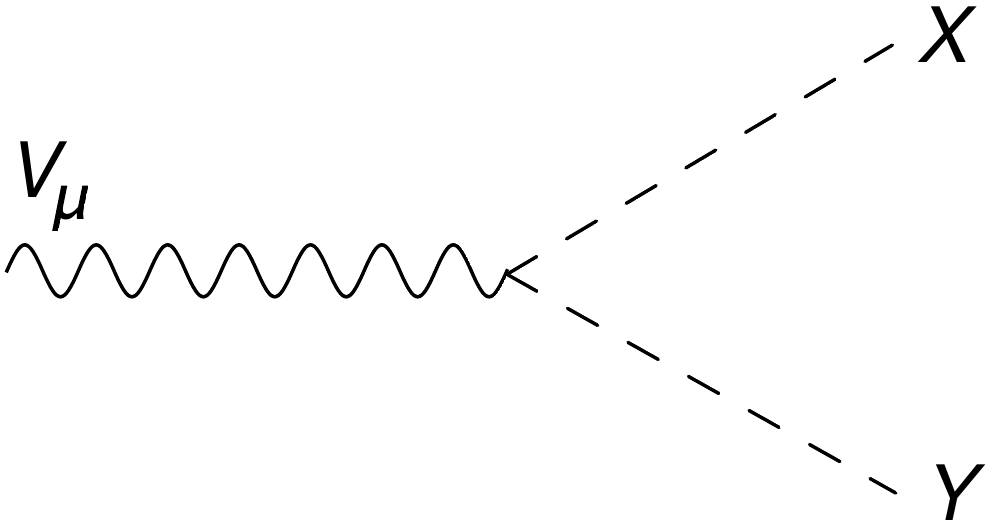}
\caption{Two scalars and a gauge boson.}\label{vxy}
\end{figure}
\begin{center}
\begin{tabular}{|c|c|}
\hline
\textbf{Interaction term} & \textbf{Feynman Rule for the vertex}\\
\hline
$\lambda V^\alpha X\partial_\alpha Y$ & $\lambda p_Y^\mu$\\
$\lambda V^\alpha X\der_\alpha Y$ & $\lambda(p_Y-p_X)^\mu$\\
\hline
\end{tabular}
\end{center}
To compute $\Pi_{ZZ}$ entering Eq.~(\ref{eq:deltarho}), we need diagrams like Fig.~\ref{vvxy} with $V=Z$. The $X,Y$ pairs are 
\begin{center}
\begin{tabular}{|c|c|c|c|}
\hline
\textbf{$X$} & \textbf{$Y$}&\textbf{Interaction term}&\textbf{Feynman Rule for the vertices}\\
\hline
$H_+$ & $H_-$        & $-\frac i2gW_3^\alpha H_+\der_\alpha H_-$ & $\frac{g^2}4(2p+q)_\mu(2p+q)_\nu$ \\
$S$   & $A_0$ & $\frac g2\frac{v_\phi}\vf W_3^\alpha S\der_\alpha A_0$ & $-\frac{g^2}4\frac{v_\phi^2}\vvf(2p+q)_\mu(2p+q)_\nu$ \\
$S$   & $a_\phi$        & $g\frac{v\sin2\beta}\vf W_3^\alpha S\partial_\alpha a_\phi$ & $-g^2\frac{v^2\sin^22\beta}{\vvf}(p+q)_\mu(p+q)_\nu$ \\
$H$   & $G_0$        & $-gW^\alpha_3H\partial_\alpha G_0$ & $-g^2(p+q)_\mu(p+q)_\nu$ \\
$G_+$ & $G_-$        & $\frac i2gW_3^\alpha G_+\der_\alpha G_- $& $\frac{g^2}4(2p+q)_\mu(2p+q)_\nu$ \\
\hline
\end{tabular}
\end{center}
To compute $\Pi_{WW}$ entering Eq.~(\ref{eq:deltarho}), we need diagrams like Fig.~\ref{vvxy} with $V=W_+$. The $X,Y$ pairs are 
\begin{center}
\begin{tabular}{|c|c|c|c|}
\hline
\textbf{$X$} & \textbf{$Y$}&\textbf{Interaction term}&\textbf{Feynman Rule for the vertices}\\
\hline
$H_+$ & $S$          & $\frac i2gW_+^\alpha H_-\der_\alpha S$ & $\frac{g^2}4(2p+q)_\mu(2p+q)_\nu$ \\
$H_+$ & $A_0$ & $\frac g2\frac{v_\phi}\vf W_+^\alpha A_0\der_\alpha H_--$ & $-\frac{g^2}4\frac{v_\phi^2}\vvf(2p+q)_\mu(2p+q)_\nu$ \\
$H_+$ & $a_\phi$        & $-g\frac{v\sin2\beta}\vf W_+^\alpha H_-\partial_\alpha a_\phi $ & $-g^2\frac{v^2\sin^22\beta}{\vvf}(p+q)_\mu(p+q)_\nu$ \\
$H$   & $G_+$        & $-gW^\alpha_+H\partial_\alpha G_-$ & $-g^2(p+q)_\mu(p+q)_\nu$ \\
$G_+$ & $G_0$        & $\frac i2gW_+^\alpha G_0\der_\alpha G_- $& $\frac{g^2}4(2p+q)_\mu(2p+q)_\nu$ \\
\hline
\end{tabular}
\end{center}


\begin{thebibliography}{99}

\bibitem{PQ} R.D. Peccei, H.R. Quinn, Phys. Rev. Lett. 38 (1977) 1440; 
S. Weinberg, Phys. Rev. Lett. 40 (1978) 223;F. Wilczek, Phys. Rev. Lett. 40 (1978) 279.

\bibitem{sik} L.Abbott and P. Sikivie, Phys. Lett. B 120, 133 (1983);
M. Dine and W. Fischler, Phys. Lett. B 120, 137 (1983);
J. Preskill, M.B. Wise and F. Wilczek, Phys. Lett. B 120, 127 (1983).

\bibitem{raff} M. Kuster, G. Raffelt and B. Beltran (eds), Lecture Notes in Physics 741 (2008).

\bibitem{dfs} M. Dine, W. Fischler and M. Srednicki, Phys. Lett. B, 104, 199 (1981).

\bibitem{models} A.R. Zhitnitsky, Sov. J. Nucl. Phys. 31, 260 (1980);
J. E. Kim, Phys. Rev. Lett. 43, 103 (1979); M. A. Shifman, A. I. Vainshtein and V. I. Zakharov, 
Nucl. Phys. B 166, 493 (1980).

\bibitem{Axion}J.~M.~Frere, J.~A.~M.~Vermaseren and M.~B.~Gavela,
  Phys.\ Lett.\ B {\bf 103}, 129 (1981);
  L.~J.~Hall and M.~B.~Wise,
  Nucl.\ Phys.\ B {\bf 187}, 397 (1981);
  H.~Georgi, D.~B.~Kaplan and L.~Randall,
  Phys.\ Lett.\ B {\bf 169}, 73 (1986);
 A.~Celis, J.~Fuentes-Martin and H.~Serodio,
  Phys.\ Lett.\ B {\bf 737}, 185 (2014)
  [arXiv:1407.0971 [hep-ph]].

\bibitem{2hdmNatural}
  R.~R.~Volkas, A.~J.~Davies and G.~C.~Joshi,
  Phys.\ Lett.\ B {\bf 215}, 133 (1988);
    R.~Foot, A.~Kobakhidze, K.~L.~McDonald and R.~R.~Volkas,
  Phys.\ Rev.\ D {\bf 89}, no. 11, 115018 (2014)
  [arXiv:1310.0223 [hep-ph]].

\bibitem{2hdmConf}
 K.~Allison, C.~T.~Hill and G.~G.~Ross,
  Phys.\ Lett.\ B
  [arXiv:1409.4029 [hep-ph]].

\bibitem{2HDM}
  G.~C.~Branco, P.~M.~Ferreira, L.~Lavoura, M.~N.~Rebelo, M.~Sher and J.~P.~Silva,
  Phys.\ Rept.\  {\bf 516}, 1 (2012)
  [arXiv:1106.0034 [hep-ph]];
    F.~J.~Botella, G.~C.~Branco, A.~Carmona, M.~Nebot, L.~Pedro and M.~N.~Rebelo,
  JHEP {\bf 1407}, 078 (2014)
  [arXiv:1401.6147 [hep-ph]];
  S.~Bertolini, L.~Di Luzio, H.~Kolesova and M.~Malinske,
  arXiv:1412.7105 [hep-ph].


\bibitem{2HDM2}
J.~F.~Gunion and H.~E.~Haber,
  Phys.\ Rev.\ D {\bf 67}, 075019 (2003)
  [hep-ph/0207010].

\bibitem{pich}
 D.~Lopez-Val, T.~Plehn and M.~Rauch,
  JHEP {\bf 1310}, 134 (2013)
  [arXiv:1308.1979 [hep-ph]];
  D.~Lopez-Val and J.~Sola,
  Phys.\ Rev.\ D {\bf 81}, 033003 (2010)
  [arXiv:0908.2898 [hep-ph]];
X.~Q.~Li, J.~Lu and A.~Pich,
  JHEP {\bf 1406}, 022 (2014)
  [arXiv:1404.5865 [hep-ph]];
A.~Celis, V.~Ilisie and A.~Pich,
  JHEP {\bf 1312}, 095 (2013)
  [arXiv:1310.7941 [hep-ph]];
 A.~Celis, V.~Ilisie and A.~Pich,
  JHEP {\bf 1307}, 053 (2013)
  [arXiv:1302.4022 [hep-ph]]; 


\bibitem{ce}  P.~Ciafaloni and D.~Espriu,
  Phys.\ Rev.\ D {\bf 56}, 1752 (1997)
  [hep-ph/9612383].

\bibitem{wudka}  A. Pomarol and R. Vega,  Nucl.Phys. B413 (1994) 3;
B. Grzadkowski, M. Maniatis and J. Wudka (UC, Riverside), JHEP 1111 (2011) 030

\bibitem{axiondecay} J. Beringer et al. (Particle Data Group), PR D86, 010001 (2012);
A.H. Corsico et al, JCAP 1212 (2012) 010;
E. Arik et al. (CAST collaboration), J. Cosmo. Astropart. Phys. 02, 008 (2009);
I. G. Irastorza et al. (IAXO collaboration), JCAP 1106 (2011) 013;
R. B{\"a}hre et al. (ALPS collaboration), JINST 1309, T09001 (2013);
S. J. Asztalos et al. (ADMX collaboration), Nuclear Instruments 
and Methods in Physics Research A 656, 39-44 (2011)


\bibitem{efcl} A.~C.~Longhitano,
  Nucl.\ Phys.\ B {\bf 188}, 118 (1981);
A.~C.~Longhitano,
  Phys.\ Rev.\ D {\bf 22}, 1166 (1980);
A. Dobado, D. Espriu and M.J. Herrero, Phys.Lett. B255 (1991) 405;
D. Espriu and M.J. Herrero, Nucl.Phys. B373 (1992) 117;
M.J. Herrero and E. Ruiz-Morales, Nucl.Phys. B418 (1994) 431; Nucl. Phys. B 437 (1995) 319; 
D. Espriu and J. Matias,  Phys.Lett. B341 (1995) 332; 

\bibitem{ey} D.~Espriu and B.~Yencho,
  Phys.\ Rev.\ D {\bf 87}, no. 5, 055017 (2013)
  [arXiv:1212.4158 [hep-ph]].

\bibitem{heff}G.~F.~Giudice, C.~Grojean, A.~Pomarol and R.~Rattazzi,
  JHEP {\bf 0706}, 045 (2007)
  [hep-ph/0703164];  
  R.~Contino, M.~Ghezzi, C.~Grojean, M.~Muhlleitner and M.~Spira,
  JHEP {\bf 1307}, 035 (2013)
  [arXiv:1303.3876 [hep-ph]];
 R.~Alonso, M.~B.~Gavela, L.~Merlo, S.~Rigolin and J.~Yepes,
  Phys.\ Lett.\ B {\bf 722}, 330 (2013)
  [arXiv:1212.3305 [hep-ph]]; 
    R.~Alonso, I.~Brivio, B.~Gavela, L.~Merlo and S.~Rigolin,
  JHEP {\bf 1412}, 034 (2014)
  [arXiv:1409.1589 [hep-ph]];
  G.~Buchalla, O.~Cat\`a and C.~Krause,
  Nucl.\ Phys.\ B {\bf 880}, 552 (2014)
  [arXiv:1307.5017 [hep-ph]]; 
  arXiv:1412.6356 [hep-ph];G.~Buchalla, O.~Cata, A.~Celis and C.~Krause,
  arXiv:1504.01707 [hep-ph]; arXiv:1511.00988 [hep-ph]; Phys.\ Lett.\ B {\bf 750}, 298 (2015) [arXiv:1504.01707 [hep-ph]],
D.~Espriu and F.~Mescia,
  Phys.\ Rev.\ D {\bf 90}, no. 1, 015035 (2014)
  [arXiv:1403.7386 [hep-ph]].
  

\bibitem{dobadollanes} R.L. Delgado, A. Dobado and F. J. Llanes-Estrada, JHEP 1402 (2014) 121; 
arXiv:1408.1193; arXiv:1502.04841;  D.~Espriu, F.~Mescia and B.~Yencho,
  Phys.\ Rev.\ D {\bf 88}, 055002 (2013)
  [arXiv:1307.2400 [hep-ph]].




\bibitem{AB} G. Altarelli and R. Barbieri, Phys. Lett. B 253, 161 (1990);
see also M. Peskin and T. Takeuchi, Phys. Rev. Lett. 65, 1964 (1990) and
A. Dobado, D. Espriu and M.J. Herrero, Phys.Lett. B255 (1991) 405; 
see Appedinx A on M.~Baak, M.~Goebel, J.~Haller, A.~Hoecker, D.~Ludwig, K.~Moenig, M.~Schott and J.~Stelzer,
  Eur.\ Phys.\ J.\ C {\bf 72}, 2003 (2012)
  [arXiv:1107.0975 [hep-ph]].



\bibitem{lhcbounds}
I. Brivio, T. Corbett, O. J. P. Eboli, M. B. Gavela, J. Gonzalez-Fraile, 
M. C. Gonzalez-Garcia,
L. Merlo and S. Rigolin, JHEP 1403, 024 (2014) [arXiv:1311.1823 [hep-ph]]; 

\bibitem{Dtexp} 
  M.~Baak, M.~Goebel, J.~Haller, A.~Hoecker, D.~Kennedy, R.~Kogler, K.~Moenig and M.~Schott {\it et al.},
  Eur.\ Phys.\ J.\ C {\bf 72}, 2205 (2012)
  [arXiv:1209.2716 [hep-ph]].


\end{thebibliography}
\end{document}